\input harvmac
%

\let\includefigures=\iftrue
\let\useblackboard=\iftrue
\newfam\black

\includefigures
\message{If you do not have epsf.tex (to include figures),}
\message{change the option at the top of the tex file.}
\input epsf
\def\figin{\epsfcheck\figin}\def\figins{\epsfcheck\figins}
\def\epsfcheck{\ifx\epsfbox\UnDeFiNeD
\message{(NO epsf.tex, FIGURES WILL BE IGNORED)}
\gdef\figin##1{\vskip2in}\gdef\figins##1{\hskip.5in}
\else\message{(FIGURES WILL BE INCLUDED)}%
\gdef\figin##1{##1}\gdef\figins##1{##1}\fi}
\def\DefWarn#1{}
\def\figinsert{\goodbreak\midinsert}
\def\ifig#1#2#3{\DefWarn#1\xdef#1{fig.~\the\figno}
\writedef{#1\leftbracket fig.\noexpand~\the\figno}%
\figinsert\figin{\centerline{#3}}\medskip\centerline{\vbox{
\baselineskip12pt\advance\hsize by -1truein
\noindent\footnotefont{\bf Fig.~\the\figno:} #2}}
\endinsert\global\advance\figno by1}
\else
\def\ifig#1#2#3{\xdef#1{fig.~\the\figno}
\writedef{#1\leftbracket fig.\noexpand~\the\figno}%
\global\advance\figno by1} \fi

\def\M{\Phi_2}
\def\tq{\tilde q}

\def\a{\alpha}
\def\la{\Lambda_1}
\def\lb{\Lambda_2}
\def\dx{\Delta x}

\def\journal#1&#2(#3){\unskip, \sl #1\ \bf #2 \rm(19#3) }
\def\andjournal#1&#2(#3){\sl #1~\bf #2 \rm (19#3) }

\def\ie{{\it i.e.}}
\def\eg{{\it e.g.}}

\noblackbox
%


\def\unlockat{\catcode`\@=11}
\def\lockat{\catcode`\@=12}

\unlockat

\def\newsec#1{\global\advance\secno by1\message{(\the\secno. #1)}
\global\subsecno=0\global\subsubsecno=0\eqnres@t\noindent
{\bf\the\secno. #1}
\writetoca{{\secsym} {#1}}\par\nobreak\medskip\nobreak}
\global\newcount\subsecno \global\subsecno=0
\def\subsec#1{\global\advance\subsecno
by1\message{(\secsym\the\subsecno. #1)}
\ifnum\lastpenalty>9000\else\bigbreak\fi\global\subsubsecno=0
\noindent{\it\secsym\the\subsecno. #1}
\writetoca{\string\quad {\secsym\the\subsecno.} {#1}}
\par\nobreak\medskip\nobreak}
\global\newcount\subsubsecno \global\subsubsecno=0
\def\subsubsec#1{\global\advance\subsubsecno by1
\message{(\secsym\the\subsecno.\the\subsubsecno. #1)}
\ifnum\lastpenalty>9000\else\bigbreak\fi
\noindent\quad{\secsym\the\subsecno.\the\subsubsecno.}{#1}
\writetoca{\string\qquad{\secsym\the\subsecno.\the\subsubsecno.}{#1}}
\par\nobreak\medskip\nobreak}

\def\subsubseclab#1{\DefWarn#1\xdef
#1{\noexpand\hyperref{}{subsubsection}%
{\secsym\the\subsecno.\the\subsubsecno}%
{\secsym\the\subsecno.\the\subsubsecno}}%
\writedef{#1\leftbracket#1}\wrlabeL{#1=#1}}
\lockat

\def\ie{{\it i.e.}}
\def\eg{{\it e.g.}}


\font\manual=manfnt \def\dbend{\lower3.5pt\hbox{\manual\char127}}

\def\IZ{\relax\ifmmode\mathchoice
{\hbox{\cmss Z\kern-.4em Z}}{\hbox{\cmss Z\kern-.4em Z}}
{\lower.9pt\hbox{\cmsss Z\kern-.4em Z}}
{\lower1.2pt\hbox{\cmsss Z\kern-.4em Z}}\else{\cmss Z\kern-.4em
Z}\fi}


\def\IZ{\relax\ifmmode\mathchoice
{\hbox{\cmss Z\kern-.4em Z}}{\hbox{\cmss Z\kern-.4em Z}}
{\lower.9pt\hbox{\cmsss Z\kern-.4em Z}}
{\lower1.2pt\hbox{\cmsss Z\kern-.4em Z}}\else{\cmss Z\kern-.4em
Z}\fi}
\def\IB{\relax{\rm I\kern-.18em B}}
\def\IC{{\relax\hbox{$\inbar\kern-.3em{\rm C}$}}}
\def\ID{\relax{\rm I\kern-.18em D}}
\def\IE{\relax{\rm I\kern-.18em E}}
\def\IF{\relax{\rm I\kern-.18em F}}
\def\IG{\relax\hbox{$\inbar\kern-.3em{\rm G}$}}
\def\IGa{\relax\hbox{${\rm I}\kern-.18em\Gamma$}}
\def\IH{\relax{\rm I\kern-.18em H}}
\def\II{\relax{\rm I\kern-.18em I}}
\def\IK{\relax{\rm I\kern-.18em K}}
\def\IP{\relax{\rm I\kern-.18em P}}
\def\IQ{\relax\hbox{$\inbar\kern-.3em{\rm Q}$}}

\def\inbar{\,\vrule height1.5ex width.4pt depth0pt}

\font\cmss=cmss10 \font\cmsss=cmss10 at 7pt
\def\IR{\relax{\rm I\kern-.18em R}}

%
%

\def\makeblankbox#1#2{\hbox{\lower\dp0\vbox{\hidehrule{#1}{#2}%
   \kern -#1
   \hbox to \wd0{\hidevrule{#1}{#2}%
      \raise\ht0\vbox to #1{}
      \lower\dp0\vtop to #1{}
      \hfil\hidevrule{#2}{#1}}%
   \kern-#1\hidehrule{#2}{#1}}}%
}%
\def\hidehrule#1#2{\kern-#1\hrule height#1 depth#2 \kern-#2}%
\def\hidevrule#1#2{\kern-#1{\dimen0=#1\advance\dimen0 by #2\vrule
    width\dimen0}\kern-#2}%
\def\openbox{\ht0=1.2mm \dp0=1.2mm \wd0=2.4mm  \raise 2.75pt
\makeblankbox {.25pt} {.25pt}  }

\def\bun#1/#2{\leavevmode
   \kern.1em \raise .5ex \hbox{\the\scriptfont0 #1}%
   \kern-.1em $/$%
   \kern-.15em \lower .25ex \hbox{\the\scriptfont0 #2}%
}

\def\opensquare{\ht0=3.4mm \dp0=3.4mm \wd0=6.8mm  \raise 2.7pt
\makeblankbox {.25pt} {.25pt}  }


\def\sector#1#2{\ {\scriptstyle #1}\hskip 1mm
\mathop{\opensquare}\limits_{\lower 1mm\hbox{$\scriptstyle#2$}}\hskip 1mm}

\def\tsector#1#2{\ {\scriptstyle #1}\hskip 1mm
\mathop{\opensquare}\limits_{\lower 1mm\hbox{$\scriptstyle#2$}}^\sim\hskip 1mm}


\def\inbar{\,\vrule height1.5ex width.4pt depth0pt}

\font\cmss=cmss10 \font\cmsss=cmss10 at 7pt
\def\IR{\relax{\rm I\kern-.18em R}}


\def\frac#1#2{{#1\over#2}}

\def\inbar{\,\vrule height1.5ex width.4pt depth0pt}
\def\IC{\relax\hbox{$\inbar\kern-.3em{\rm C}$}}
\def\IR{\relax{\rm I\kern-.18em R}}
\def\IP{\relax{\rm I\kern-.18em P}}

%
%
\catcode`\@=11
\def\slash#1{\mathord{\mathpalette\c@ncel{#1}}}
\overfullrule=0pt

\def\II{{\cal I}}

\def\SS{{\cal S}}

\def\underrel#1\over#2{\mathrel{\mathop{\kern\z@#1}\limits_{#2}}}

\catcode`\@=12


%

\def\exp{{\rm exp}}



\def\frac#1#2{{#1\over#2}}

\def\inbar{\,\vrule height1.5ex width.4pt depth0pt}
\def\IC{\relax\hbox{$\inbar\kern-.3em{\rm C}$}}
\def\IR{\relax{\rm I\kern-.18em R}}
\def\IP{\relax{\rm I\kern-.18em P}}

%
%

%
\catcode`\@=11
\def\slash#1{\mathord{\mathpalette\c@ncel{#1}}}
\overfullrule=0pt

\def\II{{\cal I}}

\def\SS{{\cal S}}

\def\underrel#1\over#2{\mathrel{\mathop{\kern\z@#1}\limits_{#2}}}

\catcode`\@=12


%

\def\exp{{\rm exp}}


\lref\GiveonSR{
  A.~Giveon and D.~Kutasov,
  ``Brane dynamics and gauge theory,''
  Rev.\ Mod.\ Phys.\  {\bf 71}, 983 (1999)
  [arXiv:hep-th/9802067].
}

\lref\AharonyTI{
  O.~Aharony, S.~S.~Gubser, J.~M.~Maldacena, H.~Ooguri and Y.~Oz,
  ``Large N field theories, string theory and gravity,''
  Phys.\ Rept.\  {\bf 323}, 183 (2000)
  [arXiv:hep-th/9905111].
}

\lref\OoguriPJ{
  H.~Ooguri and Y.~Ookouchi,
  ``Landscape of supersymmetry breaking vacua in geometrically realized gauge
  theories,''
  Nucl.\ Phys.\ B {\bf 755}, 239 (2006)
  [arXiv:hep-th/0606061].
}

\lref\ArgurioNY{
  R.~Argurio, M.~Bertolini, S.~Franco and S.~Kachru,
  ``Gauge / gravity duality and meta-stable dynamical supersymmetry breaking,''
  arXiv:hep-th/0610212.
}

\lref\AganagicEX{
  M.~Aganagic, C.~Beem, J.~Seo and C.~Vafa,
  ``Geometrically induced metastability and holography,''
  arXiv:hep-th/0610249.
}

\lref\HeckmanWK{
  J.~J.~Heckman, J.~Seo and C.~Vafa,
  ``Phase Structure of a Brane/Anti-Brane System at Large N,''
  arXiv:hep-th/0702077.
}

\lref\OoguriWJ{
  H.~Ooguri and C.~Vafa,
  ``Two-Dimensional Black Hole and Singularities of CY Manifolds,''
  Nucl.\ Phys.\ B {\bf 463}, 55 (1996)
  [arXiv:hep-th/9511164].
}

\lref\KutasovTE{
  D.~Kutasov,
  ``Orbifolds and Solitons,''
  Phys.\ Lett.\ B {\bf 383}, 48 (1996)
  [arXiv:hep-th/9512145].
}

\lref\GiveonZM{
  A.~Giveon, D.~Kutasov and O.~Pelc,
  ``Holography for non-critical superstrings,''
  JHEP {\bf 9910}, 035 (1999)
  [arXiv:hep-th/9907178].
}

\lref\OoguriBG{
  H.~Ooguri and Y.~Ookouchi,
  ``Meta-stable supersymmetry breaking vacua on intersecting branes,''
  Phys.\ Lett.\ B {\bf 641}, 323 (2006)
  [arXiv:hep-th/0607183].
}

\lref\FrancoHT{
  S.~Franco, I.~Garcia-Etxebarria and A.~M.~Uranga,
  ``Non-supersymmetric meta-stable vacua from brane configurations,''
  arXiv:hep-th/0607218.
}

\lref\BenaRG{
  I.~Bena, E.~Gorbatov, S.~Hellerman, N.~Seiberg and D.~Shih,
  ``A note on (meta)stable brane configurations in MQCD,''
  JHEP {\bf 0611}, 088 (2006)
  [arXiv:hep-th/0608157].
}

\lref\AhnGN{
  C.~Ahn,
  ``Brane configurations for nonsupersymmetric meta-stable vacua in SQCD with
  adjoint matter,''
  arXiv:hep-th/0608160.
}

\lref\TatarDM{
  R.~Tatar and B.~Wetenhall,
  ``Metastable vacua, geometrical engineering and MQCD transitions,''
  arXiv:hep-th/0611303.
}

\lref\IntriligatorDD{
  K.~Intriligator, N.~Seiberg and D.~Shih,
  ``Dynamical SUSY breaking in meta-stable vacua,''
  JHEP {\bf 0604}, 021 (2006)
  [arXiv:hep-th/0602239].
}

\lref\SakaiCN{
  T.~Sakai and S.~Sugimoto,
  ``Low energy hadron physics in holographic QCD,''
  Prog.\ Theor.\ Phys.\  {\bf 113}, 843 (2005)
  [arXiv:hep-th/0412141].
}

\lref\AntonyanVW{
  E.~Antonyan, J.~A.~Harvey, S.~Jensen and D.~Kutasov,
  ``NJL and QCD from string theory,''
  arXiv:hep-th/0604017.
}

\lref\AntonyanQY{
  E.~Antonyan, J.~A.~Harvey and D.~Kutasov,
  ``The Gross-Neveu model from string theory,''
  arXiv:hep-th/0608149.
}

\lref\AntonyanPG{
  E.~Antonyan, J.~A.~Harvey and D.~Kutasov,
  ``Chiral symmetry breaking from intersecting D-branes,''
  arXiv:hep-th/0608177.
}

\lref\CallanAT{
  C.~G.~Callan, J.~A.~Harvey and A.~Strominger,
  ``Supersymmetric string solitons,''
  arXiv:hep-th/9112030.
}

\lref\AharonyUB{
  O.~Aharony, M.~Berkooz, D.~Kutasov and N.~Seiberg,
  ``Linear dilatons, NS5-branes and holography,''
  JHEP {\bf 9810}, 004 (1998)
  [arXiv:hep-th/9808149].
}

\lref\AharonyXN{
  O.~Aharony, A.~Giveon and D.~Kutasov,
  ``LSZ in LST,''
  Nucl.\ Phys.\ B {\bf 691}, 3 (2004)
  [arXiv:hep-th/0404016].
}

\lref\ElitzurFH{
  S.~Elitzur, A.~Giveon and D.~Kutasov,
  ``Branes and N = 1 duality in string theory,''
  Phys.\ Lett.\ B {\bf 400}, 269 (1997)
  [arXiv:hep-th/9702014].
}

\lref\ElitzurHC{
  S.~Elitzur, A.~Giveon, D.~Kutasov, E.~Rabinovici and A.~Schwimmer,
  ``Brane dynamics and N = 1 supersymmetric gauge theory,''
  Nucl.\ Phys.\ B {\bf 505}, 202 (1997)
  [arXiv:hep-th/9704104].
}

\lref\SeibergPQ{
  N.~Seiberg,
  ``Electric - magnetic duality in supersymmetric nonAbelian gauge theories,''
  Nucl.\ Phys.\ B {\bf 435}, 129 (1995)
  [arXiv:hep-th/9411149].
}

\lref\KutasovDJ{
  D.~Kutasov,
  ``D-brane dynamics near NS5-branes,''
  arXiv:hep-th/0405058.
}

\lref\KitanoXG{
  R.~Kitano, H.~Ooguri and Y.~Ookouchi,
  ``Direct mediation of meta-stable supersymmetry breaking,''
  arXiv:hep-ph/0612139.
}

\lref\KutasovRR{
  D.~Kutasov,
  ``Accelerating branes and the string / black hole transition,''
  arXiv:hep-th/0509170.
}

\lref\ItzhakiZR{
  N.~Itzhaki, D.~Kutasov and N.~Seiberg,
  ``Non-supersymmetric deformations of non-critical superstrings,''
  JHEP {\bf 0512}, 035 (2005)
  [arXiv:hep-th/0510087].
}

\lref\LukyanovNJ{
  S.~L.~Lukyanov, E.~S.~Vitchev and A.~B.~Zamolodchikov,
  ``Integrable model of boundary interaction: The paperclip,''
  Nucl.\ Phys.\ B {\bf 683}, 423 (2004)
  [arXiv:hep-th/0312168].
}

\lref\LukyanovBF{
  S.~L.~Lukyanov and A.~B.~Zamolodchikov,
  ``Dual form of the paperclip model,''
  Nucl.\ Phys.\ B {\bf 744}, 295 (2006)
  [arXiv:hep-th/0510145].
}

\lref\KutasovCT{
  D.~Kutasov,
  ``A geometric interpretation of the open string tachyon,''
  arXiv:hep-th/0408073.
}

\lref\GiveonPR{
  A.~Giveon and D.~Kutasov,
  ``Fundamental strings and black holes,''
  arXiv:hep-th/0611062.
}

\lref\WittenSC{
  E.~Witten,
  ``Solutions of four-dimensional field theories via M-theory,''
  Nucl.\ Phys.\  B {\bf 500}, 3 (1997)
  [arXiv:hep-th/9703166].
}

\lref\AdamsSV{
  A.~Adams, J.~Polchinski and E.~Silverstein,
  ``Don't panic! Closed string tachyons in ALE space-times,''
  JHEP {\bf 0110}, 029 (2001)
  [arXiv:hep-th/0108075].
}

\lref\HarveyWM{
  J.~A.~Harvey, D.~Kutasov, E.~J.~Martinec and G.~W.~Moore,
  ``Localized tachyons and RG flows,''
  arXiv:hep-th/0111154.
}

\lref\NakayamaYX{
Y.~Nakayama, Y.~Sugawara and H.~Takayanagi,
``Boundary states for the rolling D-branes in NS5 background,''
JHEP {\bf 0407}, 020 (2004)
[arXiv:hep-th/0406173].
}

\lref\HananyIE{
A.~Hanany and E.~Witten,
``Type IIB superstrings, BPS monopoles, and three-dimensional gauge
dynamics,''
Nucl.\ Phys.\  B {\bf 492}, 152 (1997)
[arXiv:hep-th/9611230].
}

\Title{
} {\vbox{ \centerline{Gauge Symmetry and Supersymmetry Breaking}
\bigskip
\centerline{From Intersecting Branes} }}
\medskip
\centerline{\it Amit Giveon${}^{1}$ and David Kutasov${}^{2}$}
\bigskip
\smallskip
\centerline{${}^{1}$Racah Institute of Physics, The Hebrew
University} \centerline{Jerusalem 91904, Israel}
\smallskip
\centerline{${}^2$EFI and Department of Physics, University of
Chicago} \centerline{5640 S. Ellis Av., Chicago, IL 60637, USA }

\bigskip\bigskip\bigskip
\noindent

We study a system of intersecting NS and D-branes in type IIA
string theory in $\IR^{9,1}$. We show that the $3+1$ dimensional
non-supersymmetric theory at the intersection has unstable vacua
which are long-lived in some regions of the parameter space of
brane configurations, and disappear in others. We also comment
on the relation of our construction to systems of $D$ and
$\bar D$-branes wrapped around cycles of non-compact
Calabi-Yau manifolds and to other related systems.

\vglue .3cm
\bigskip

\Date{3/07}

\bigskip

\newsec{Introduction}

Much of the work on brane dynamics in string theory in the past
decade (see \eg\ \refs{\GiveonSR,\AharonyTI} for reviews) focused on
supersymmetric backgrounds, which are typically easier to control
than generic non-supersymmetric ones. Extending the results to
non-supersymmetric backgrounds is important for a number of reasons.
First, they are expected to exhibit a richer set of dynamical
phenomena. Second, their study might help to address the vacuum
selection problem in string theory. Finally, they may be useful for
constructing a model of nature below the scale of supersymmetry
breaking.

In the supersymmetric case two main classes of constructions were
considered in the past. One involves D-branes in the vicinity of
regular or singular points on Calabi-Yau (CY) manifolds. D-branes
localized near such points give rise to gauge theories some of whose
properties can be analyzed using the geometric realization. One can
take the CY manifold to be non-compact so that the four dimensional
Newton constant vanishes, since gravity is secondary for the
analysis. Of course, in order to use such constructions in realistic
compactifications, they eventually have to be embedded in a compact
CY.

One way to break supersymmetry  in this framework is to place at a
CY singularity D-branes which do not preserve any supersymmetry.
Since the theory without the D-branes is supersymmetric, any
instabilities associated with the lack of supersymmetry are
typically relatively mild and can be resolved by rearrangement or
annihilation of the branes, as is familiar from open string tachyon
condensation. Recent discussions of non-supersymmetric  D-brane
systems on CY manifolds include
\refs{\OoguriPJ\ArgurioNY\AganagicEX-\HeckmanWK}.

A second class of constructions (reviewed in  \GiveonSR) involves D-branes
in the vicinity of Neveu-Schwarz (NS) fivebranes. These constructions are
closely related to the previous ones, with the fivebranes serving as an analog
of the CY geometry.  When some of the directions transverse to the fivebranes
are compact, the two types of constructions are related by T-duality
\refs{\OoguriWJ\KutasovTE-\GiveonZM}. Again, if one is not
interested in gravitational physics, all directions transverse to
the intersection of the branes can be taken to be non-compact.

In this note we will analyze some non-supersymmetric systems of
intersecting D-branes and $NS5$-branes. We will focus on systems
with $3+1$ dimensional intersections and consider situations in
which the $NS5$-branes alone preserve some supersymmetry (as in
the CY case). Adding the D-branes breaks supersymmetry and leads to
potential instabilities. We would like to determine the resulting low
energy dynamics as a function of the parameters of the brane configurations.

As is familiar from other brane systems, in different regions of
this parameter space one can study the dynamics using
different tools. In one region the correct description is in terms
of a low energy gauge theory of the sort analyzed in
\IntriligatorDD.\foot{The brane configuration corresponding to the
gauge theory of \IntriligatorDD\ was studied in
\refs{\OoguriBG\FrancoHT\BenaRG\AhnGN-\TatarDM};
we will comment on it below.} In another, one can use a
Dirac-Born-Infeld (DBI) action for the D-branes in the background of
the fivebranes. The resulting analysis is very similar to the generalization
of the Sakai-Sugimoto \SakaiCN\ model of dynamical chiral symmetry breaking
studied in \refs{\AntonyanVW\AntonyanQY-\AntonyanPG}. The main difference is
that there the symmetry that is broken by the dynamics is global, while here
it is a gauge symmetry. At the same time, many of the techniques that were
used there can be applied here.

The plan of this paper is the following. In section 2 we introduce
the brane configuration that is going to serve as our primary
example and briefly discuss some of its properties. We show that
already at the most elementary level of consideration, as one changes
the parameters of this system, it undergoes a first order phase
transition at which the nature of the vacuum changes significantly.
We also discuss the gauge theory that corresponds to it at low energies.

In section 3 we study the phase structure of this system more precisely,
by including the gravitational interaction between the $NS$ and
$D$-branes. This interaction modifies the dynamics at string tree
level and can be studied by analyzing the DBI action for the $D$-branes
in the background of the fivebranes. We find a rich structure that
depends on the parameters of the brane configuration. In one region
in parameter space we find a first order phase transition; in another,
a second order one. We also comment on the corrections to the DBI
approximation due to string $(\alpha')$ effects.

In section 4 we briefly describe the quantum dynamics of our brane configuration
in a regime in parameter space where it is well described by a low energy
gauge theory. The results of \IntriligatorDD\ imply that in this region
supersymmetry, which is broken classically, is restored quantum mechanically,
and the classical states discussed in section 3 are meta-stable. In section 5
we discuss a possible extrapolation of this picture to the regime discussed in
section 3, where the gauge theory analysis is unreliable.

In section 6 we discuss our results and possible directions for further research.
We also comment on the relation of our system to the brane construction relevant
for the gauge theory of \IntriligatorDD, and to the systems discussed in
\refs{\AganagicEX,\HeckmanWK}, which contain branes and antibranes wrapped around
different cycles of CY manifolds.

\newsec{A type IIA brane configuration}

\subsec{Fivebranes}

The starting point of our discussion is a supersymmetric
configuration of $NS5$-branes, to which we will later add D-branes
that break supersymmetry. All the branes will be extended in the
$\IR^{3,1}$ labeled by $(0123)$, the physical spacetime which will
serve as the arena for the dynamics of interest. We will use two
types of $NS5$-branes, which we will refer to as $NS$ and
$NS'$-branes (as in \GiveonSR) and are oriented as follows:
\eqn\orientfive{\eqalign{ NS:&\qquad\qquad(012345)\cr
                          NS':&\qquad\qquad(012389)\cr
}}
It is easy to check that the branes \orientfive\ preserve $N=2$
supersymmetry in $3+1$ dimensions (see \eg\ \GiveonSR). Since the
directions $(56)$ will play a special role below, we will introduce
special notation for them,
\eqn\notfs{(x^5,x^6)=(x,y)~.}
The fivebrane configuration of interest is depicted in figure 1. It
includes $k$ coincident $NS$-branes extended in $x$ and localized at
$y=0$, and two $NS'$-branes, $NS'_1$ and $NS'_2$, at
$(x,y)=(x_1,y_1)$ and $(x_2,y_2)$, respectively.

\ifig\loc{The configuration of $NS$ and $NS'$-branes.}
{\epsfxsize3.0in\epsfbox{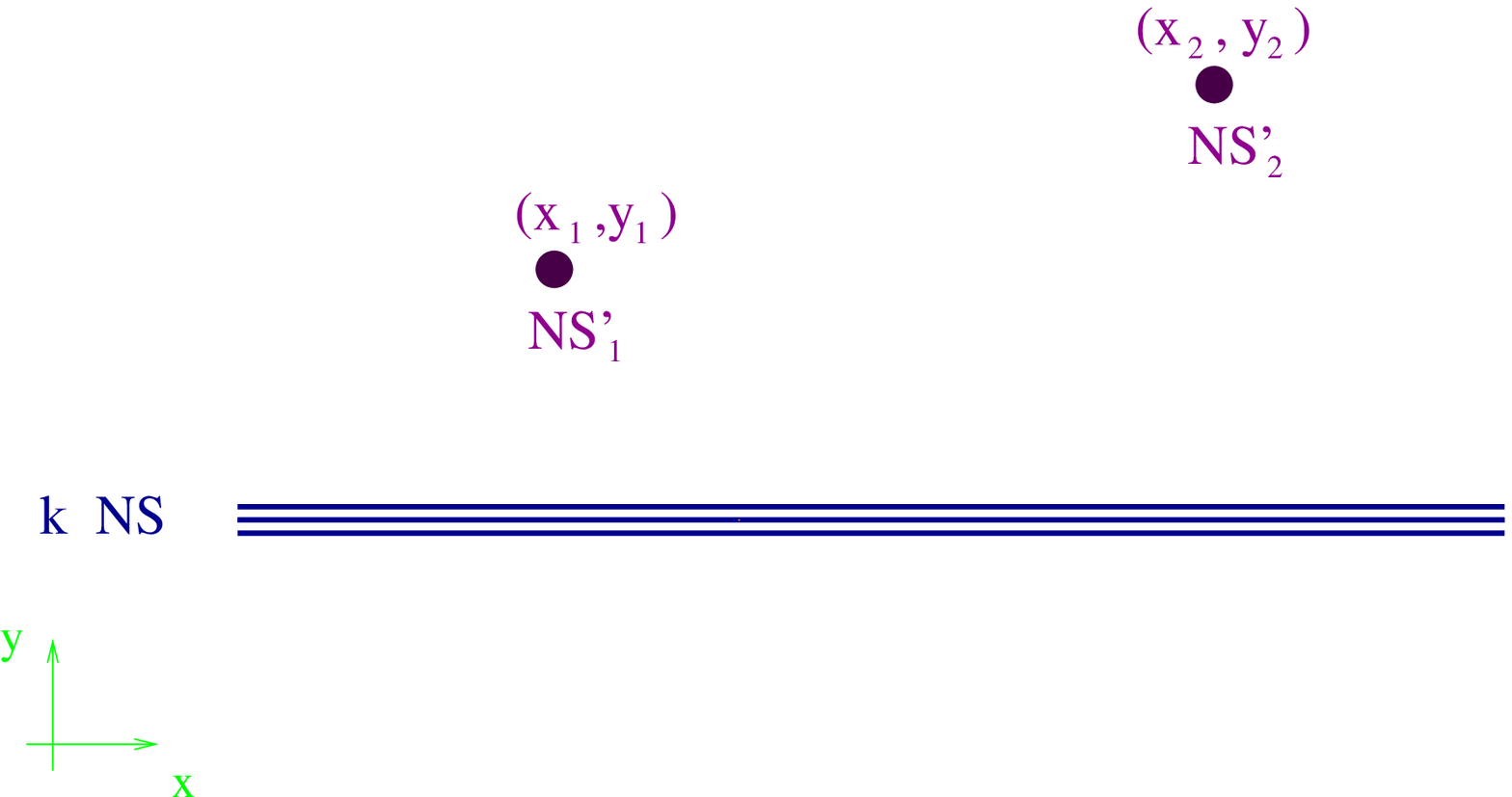}}

For $k=1$, each $(NS,NS')$ intersection can be thought of as dual
to a conifold singularity on a non-compact CY manifold.
This description is particularly useful in the limit $y_i\ll l_s$ in
which the geometry near each intersection develops a long throat.
For large $\Delta x=x_2-x_1$, the configuration of figure 1 can be
thought of as describing two widely separated conifolds. As $\Delta
x$ decreases, the two conifolds approach each other and the geometry
becomes more complicated. It can be studied using the techniques of
\GiveonZM.

The $k$ coincident $NS$ branes in figure 1 curve the geometry at the
characteristic scale
\eqn\defll{l=\sqrt{k}l_s}
Thus, for large $k$ the curving of the geometry is primarily due to the
$NS$-branes. This is one of the reasons to introduce $k$ in the first place.
We will see later that the important dynamics here is in any case due to
the effect of the $NS$-branes so that many of the results below are valid
for small $k$ as well.

The geometry of the $k$ $NS$-branes in figure 1 is given by
\eqn\chs{\eqalign{ &ds^2=dx_\mu dx^\mu+H(x^n) dx_mdx^m\cr
&e^{2(\Phi-\Phi_0)}=H(x^n)\cr
&H_{mnp}=-\epsilon_{mnp}^q\partial_q\Phi~.\cr }}
Here $\mu=0,1,2,3,4,5$; $m=6,7,8,9$; $H_{mnp}$ is the field strength
of the Neveu-Schwarz $B$ field; $g_s=\exp\Phi_0$ is the string
coupling far from the fivebranes. The harmonic function $H$ is given
by
\eqn\hcoin{H(r)=1+{kl_s^2\over r^2}=1+{l^2\over r^2}~,}
with $r^2=x_mx^m$. This geometry was found by C. Callan, J. Harvey
and A. Strominger \CallanAT\ by solving the low energy equations of
motion of string theory, but these authors pointed out that due to
its high degree of worldsheet supersymmetry it should not receive
$\alpha'$ corrections. Subsequent studies in the context of Little
String Theory (see \eg\ \refs{\AharonyUB,\AharonyXN}) have provided
further support for this expectation. Thus, we will take the point
of view that we can use it as long as the local string
coupling is small. This is the case everywhere except in a
region of (approximate) size $g_sl$ around the fivebranes.

\subsec{Adding $D4$ and $\overline{D4}$-branes}

As mentioned in the introduction, a natural way to break
supersymmetry is to add to the fivebrane configuration described
in the previous subsection D-branes that break supersymmetry. For
example, one can add $N_2$ $D4$-branes stretched between the
$NS$-branes and the $NS'_2$-brane, and $N_1$
$\overline{D4}$-branes stretched between the $NS$-branes and the
$NS'_1$-brane (see figure 2). We will take $N_2\ge N_1$ below.

\ifig\loc{The brane configuration with $D4$ and
$\overline{D4}$-branes (shown in red).}
{\epsfxsize3.0in\epsfbox{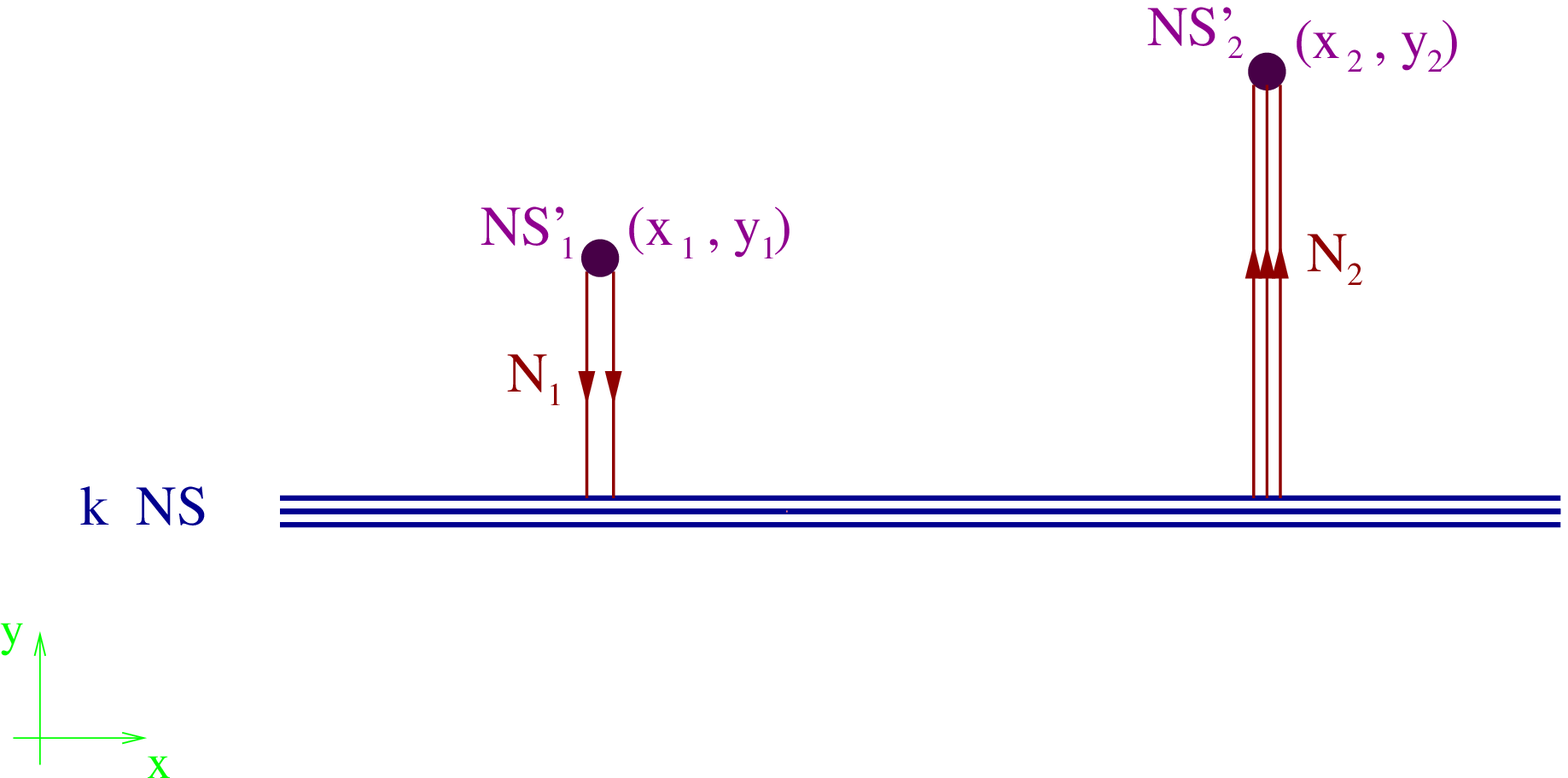}}

Note that the $D4$ and $\overline{D4}$-branes in figure 2 cannot
annihilate, since they carry different charges under the gauge
symmetries on the two $NS'$-branes. This is different from systems
such as that discussed in  \refs{\AganagicEX,\HeckmanWK}, which
contains two conifold singularities some distance apart, with
$D$-branes and $\overline{D}$-branes wrapping small spheres associated
with the two conifolds, respectively. In that system the branes and
anti-branes can annihilate after overcoming a potential barrier. We
will comment on it further in section 6.

The brane configuration of figure 2 contains the free parameters
$(x_i,y_i)$ $i=1,2$ (which are naturally measured in units of $l$,
\defll), and the string coupling $g_s$. In this and the next section
we will study it in the (semi-) classical approximation, \ie\ keep
$(x_i,y_i)$ fixed and send the string coupling to zero. Our purpose
will be to analyze the low energy dynamics in this limit. In sections
4, 5 we will discuss $g_s$ corrections.

The configuration of figure 2 is stable when the two $NS'$-branes
are sufficiently far apart, but if they are close it is unstable to
decay to the configuration of figure 3.

\ifig\loc{A brane configuration in the same charge sector as that of
figure 2.} {\epsfxsize3.0in\epsfbox{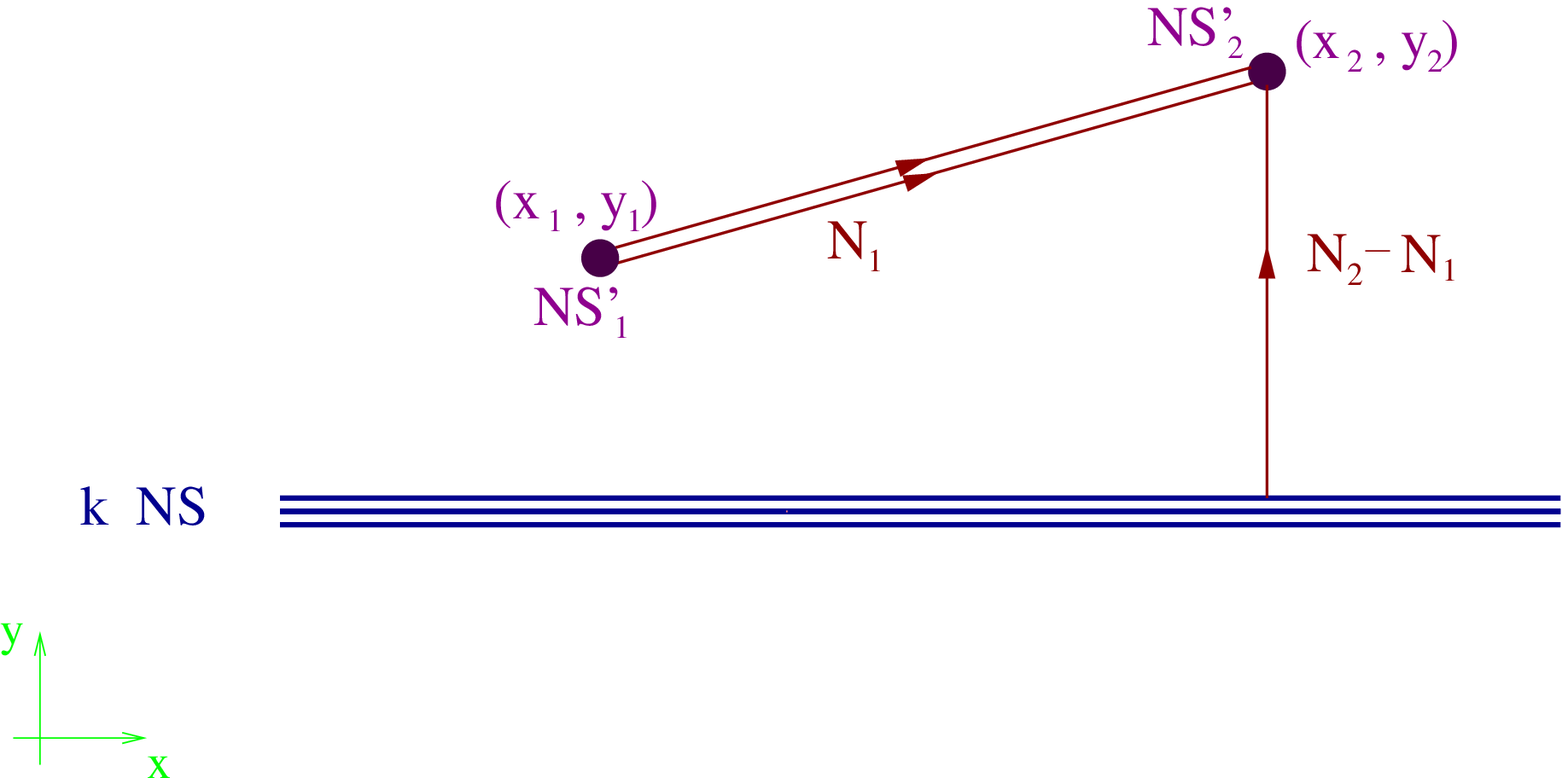}}

\noindent
Indeed, the difference between the energy densities of the
configurations of figures 3 and 2 is ($\tau_4$ is the tension of a
$D4$-brane)
\eqn\diffen{V_3-V_2=N_1\tau_4\left[\sqrt{\Delta x^2+\Delta
y^2}-(y_1+y_2)\right]~.}
This is positive for
\eqn\posneg{\Delta x>2\sqrt{y_1y_2}~,}
and negative otherwise. Thus, for $\Delta x$ in the range \posneg\
the ground state of the system is the configuration of figure 2;
otherwise it is that of figure 3. The transition between the two, 
at $\Delta x=2\sqrt{y_1y_2}$, is a first order phase transition. 
Indeed, as is clear from figures 2, 3, the ground state (and in particular
the gauge group and massless matter content) changes abruptly 
across the transition.

While for $\Delta x<2\sqrt{y_1y_2}$ the configuration of figure 2 is
not the lowest energy one, it is still locally stable. In order to
make the transition to the configuration of figure 3, the ends of
$N_1$ $D4$ and $\overline{D4}$-branes on the $NS$-branes have to meet
and reconnect. In the process, the energy of the $D$-branes
increases before decreasing to that of the configuration of figure
3. Thus, in this regime the configuration of figure 2 is meta-stable
(for small $g_s$). Naively this is true for arbitrarily small $\Delta x$
(larger than the string length), but we will see later that a more careful
analysis leads to different conclusions.

Similarly, for $\Delta x$ in the range \posneg\ the configuration of
figure 3 is locally stable and is separated by a potential barrier
from the actual ground state, which is the configuration in figure
2. Naively this is true for arbitrarily large $\Delta x$, but we will
see later that a more careful analysis leads to different conclusions.

\subsec{Gauge theory on the $D4$-branes}

In a certain region in the parameter space of the brane
configuration of figure 2, the low energy dynamics is well described
by the gauge theory on the $D4$-branes. This gauge theory can be
constructed as follows. Imagine starting with the configuration of
figure 2 and taking the $NS'_1$-brane around the $NS$-branes (in the
$x^7$ direction) without varying the length of the $D4$-branes
connecting the two. At some point all the $D4$-branes in figure 2
align, as in figure 4a, and supersymmetry is restored.

\ifig\loc{The brane configuration corresponding to the electric gauge
theory, with non-vanishing (a), and vanishing (b) mass for the
bifundamentals.}{\epsfxsize5.0in\epsfbox{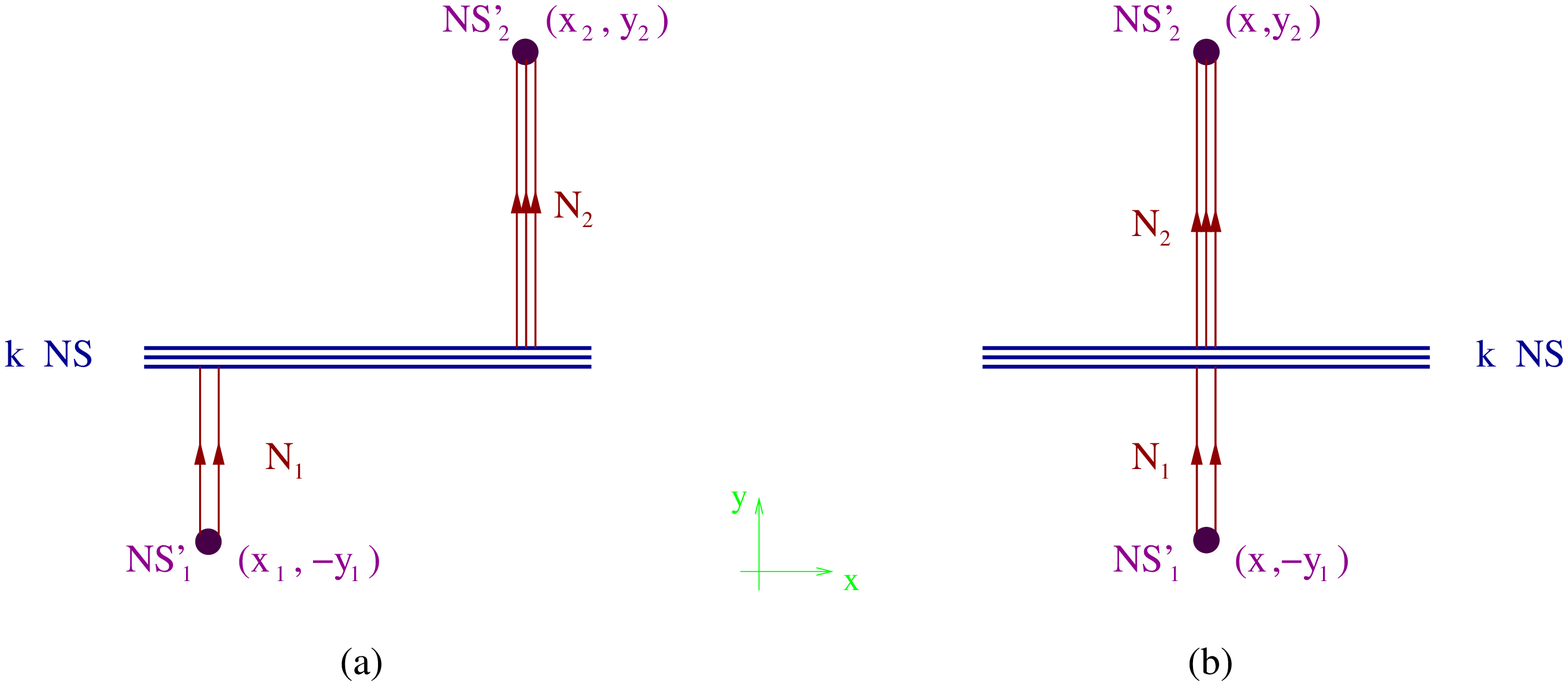}}

If one further takes $\Delta x=x_2-x_1\to 0$, one arrives at the
configuration of figure 4b.
For $k=1$ this configuration corresponds to an $N=1$ supersymmetric
gauge theory with gauge group
\eqn\gggg{U(N_1)\times U(N_2)~,}
and bifundamental chiral superfields $Q$, $\tilde Q$. For $k>1$
there is also an adjoint of the gauge group \gggg, $\Phi$, with a
polynomial superpotential \GiveonSR.

The gauge couplings of $U(N_1)$ and $U(N_2)$, $g_1$ and $g_2$, are
given as usual by
\eqn\gaugeonetwo{g_i^2={g_s l_s\over y_i}~.}
In particular, by varying the distances $y_i$ one can change the
relative strength of the two couplings. E.g. in the limit
$y_2\to\infty$, $U(N_2)$ becomes a global symmetry and the theory
becomes SQCD with gauge group $U(N_1)$ and $N_2$ flavors in the
fundamental representation.

As is standard in constructions of this sort \GiveonSR, deformations
of the brane configuration of figure 4b correspond to parameters in
the gauge theory \gggg. For example, displacing the two $NS'$-branes
relative to each other in the $x$ direction by an amount $\Delta x$,
as in figure 4a, corresponds to turning on a quadratic superpotential
for the bifundamentals,
\eqn\massqq{W=m\tilde Q Q\equiv mM~,}
where $M$ is the meson field, and the mass $m$ is given by
\eqn\formm{m={\Delta x\over2\pi\alpha'}~.}
Relative displacements between the $NS'$-branes and the $NS$-brane
in $x^7$ correspond to Fayet-Iliopoulos D-terms in the $U(1)$
factors of the gauge group \gggg. Reconnecting some of the
$D4$-branes in figure 4b such that they stretch between the two
$NS'$-branes and moving them off in the $(89)$ directions
corresponds to giving an expectation value to the meson field $M$
along one of its flat directions.

{}From the point of view of the brane configuration of figure 4,
configurations in which the two $NS'$-branes are on the same side of
the $NS$-brane, as in figures 2, 3, are obtained
\refs{\ElitzurFH,\ElitzurHC} by applying Seiberg duality to the
$U(N_1)$ factor in \gggg. Indeed, starting with figure 4b and
performing the opposite move to the one that led us to it, namely
taking the $NS'_1$-brane around the $NS$-brane in $x^7$ until it
hits the $D4$-branes stretched between the $NS$ and $NS'_2$-branes,
leads\foot{After the annihilation of $N_1$ branes and antibranes by
open string tachyon condensation.} to the configuration of figure 5.

\ifig\loc{The brane configuration corresponding to the magnetic
gauge theory.} {\epsfxsize3.0in\epsfbox{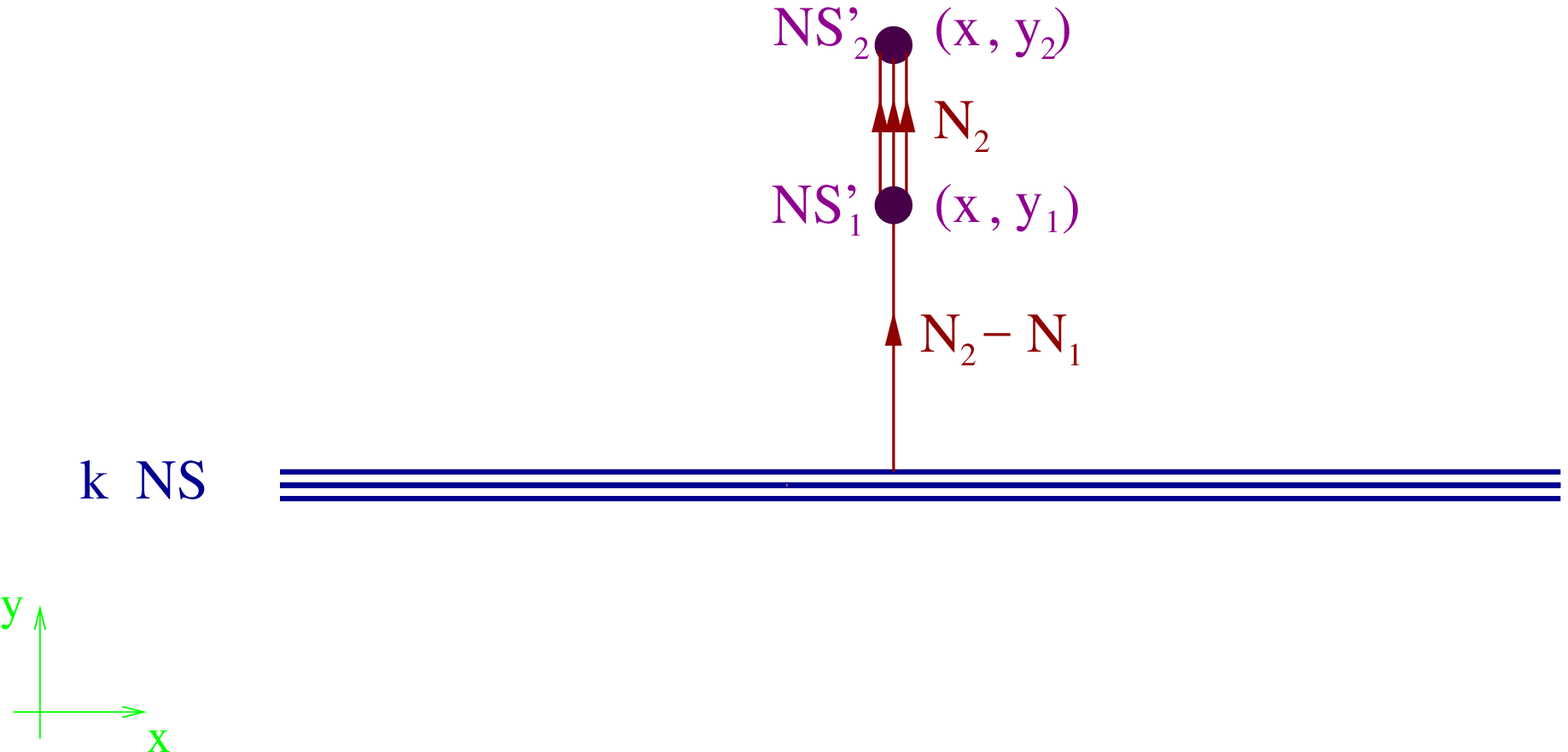}}

The corresponding gauge theory has gauge group
\eqn\maggg{U(N_2-N_1)\times U(N_2)~,}
bifundamental ``quarks'' $q$, $\tilde q$, and an adjoint of
$U(N_2)$, $M$ (related to the electric mesons \massqq), coupled to
the quarks via a cubic superpotential
\eqn\cubsup{W_{\rm mag}={1\over\Lambda}M\tq q~,}
where $\Lambda$ is an energy scale. This is precisely the matter
content and interactions of the Seiberg dual \SeibergPQ\ of \gggg.
Configurations such as those of figures 2,3 are obtained by turning
on the superpotential \massqq, which in the magnetic language
corresponds to adding a term linear in $M$ to the superpotential  
\cubsup,
\eqn\linsup{W_{\rm mag}={1\over\Lambda}M\tq q+mM~.}
One can of course go directly from the brane configuration of figure 2
to that of figure 5 without passing through the ``electric'' configurations
of figure 4. Indeed, taking the separation $\Delta x$ (or equivalently, \formm,
the parameter $m$ in \linsup) in figure 2 to zero leads precisely to figure 5.

Conversely, one can think of the brane configurations of figures 2, 3 as particular
states in the magnetic gauge theory with gauge group \maggg\ and superpotential
\linsup. Of course, the gauge theory description is only valid at low energies.
This means that the mass parameter $m$ and thus the separation of $NS'$-branes
$\Delta x$ \formm\ must be sufficiently small for this description to be valid.
If $m$ is too large, the relevant description is not the gauge theory one, but
rather that of the UV completion of the theory via brane dynamics in string theory.

We will return to the gauge theory described above in section 4. We will see
that it is closely related to the theory studied in \IntriligatorDD, and in
particular has meta-stable supersymmetry breaking vacua. Our main motivation is
to understand how the analysis of \IntriligatorDD\ ties in with the behavior
of the brane system under consideration in other regimes in parameter space,
where one needs to use other techniques for studying it.

\newsec{DBI analysis}

In section 2 we discussed the brane configurations of figures 2 -- 5
in the classical, flat space limit, in which the fivebranes are
treated as hypersurfaces on which D-branes can end. For a more
accurate description one needs to take into account certain classical
and quantum $(g_s)$ corrections. In this section we will
discuss the former; the latter will be discussed later in the paper.

\subsec{General analysis}

The main source of classical corrections is the gravitational
potential created by the $k$ $NS$-branes \chs, \hcoin, and its
effects on the $D4$-branes. One reason for including it in the
analysis has to do with the transition between the configurations of
figures 2 and 3 which, as discussed in section 2, occurs as we vary
the parameters of the brane configuration. This transition takes
place near the $NS$-branes where the geometry \chs\ is nontrivial.
Another reason is that while the discussion of section 2.2 might be
expected to be reliable for large separations of the
branes\foot{Although we will see that even in that regime the
fivebrane geometry gives rise to important new qualitative
effects.}, the fivebrane geometry gives large corrections when the
parameters $y_i$ in figures 2, 3 are comparable to $l$ \defll.

We will describe the D-branes by a DBI action in the $NS$-brane
backround \chs. This description is reliable for large $k$ but
some aspects of it are valid for all $k$. We will comment on this
issue later in the section.

The straight brane configuration of figure 2 is a solution of the
DBI equations of motion for all values of $l$. In the configuration
of figure 3, the geometry \chs\ gives rise to an attractive force
that pulls the $N_1$ $D4$-branes stretched between the $NS'$-branes
towards the $NS$-branes. To calculate their shape one needs to
analyze the DBI action for $D4$-branes in the geometry \chs\ (see
\KutasovDJ\ for a related discussion).

We are looking for a solution in which the $D4$-branes are described
by a smooth curve $y=y(x)$ connecting the points $(x_1,y_1)$ and
$(x_2, y_2)$. Its shape is obtained by extremizing the DBI action
\eqn\dbiradial{\SS_4=-N_1\tau_4\int dx{1\over\sqrt{H(y)}}
\sqrt{1+H(y)(\partial_x y)^2}= -N_1\tau_4\int dx \sqrt{ {1\over
H(y)}+(\partial_x y)^2} ~,}
where $H(y)$ is given by \hcoin. The equations of motion of this
action have a first integral
\eqn\firstint{H(y)\sqrt{{1\over H(y)}+(\partial_x y)^2}=C~.}
To solve \firstint\ it is useful to think about the qualitative
form of the solution as a function of
\eqn\ddxx{\Delta x=x_2-x_1~.}
{}For $\Delta x=0$, figure 3 reduces to figure 5 and the $N_1$
$D4$-branes connecting the $NS'$-branes stretch vertically, along
the $y$ direction. For non-zero $\Delta x$, the solution is a
deformation of the $N_1$ straight branes in figure 3. If $\Delta x$
is small enough, $y$ is a monotonic function of $x$ everywhere. This
regime can be studied using the methods below, but we will not
describe it in detail here.

\ifig\loc{The effect of the gravitational attraction to the
$NS$-branes on the $D4$-branes in figure 3.}
{\epsfxsize3.0in\epsfbox{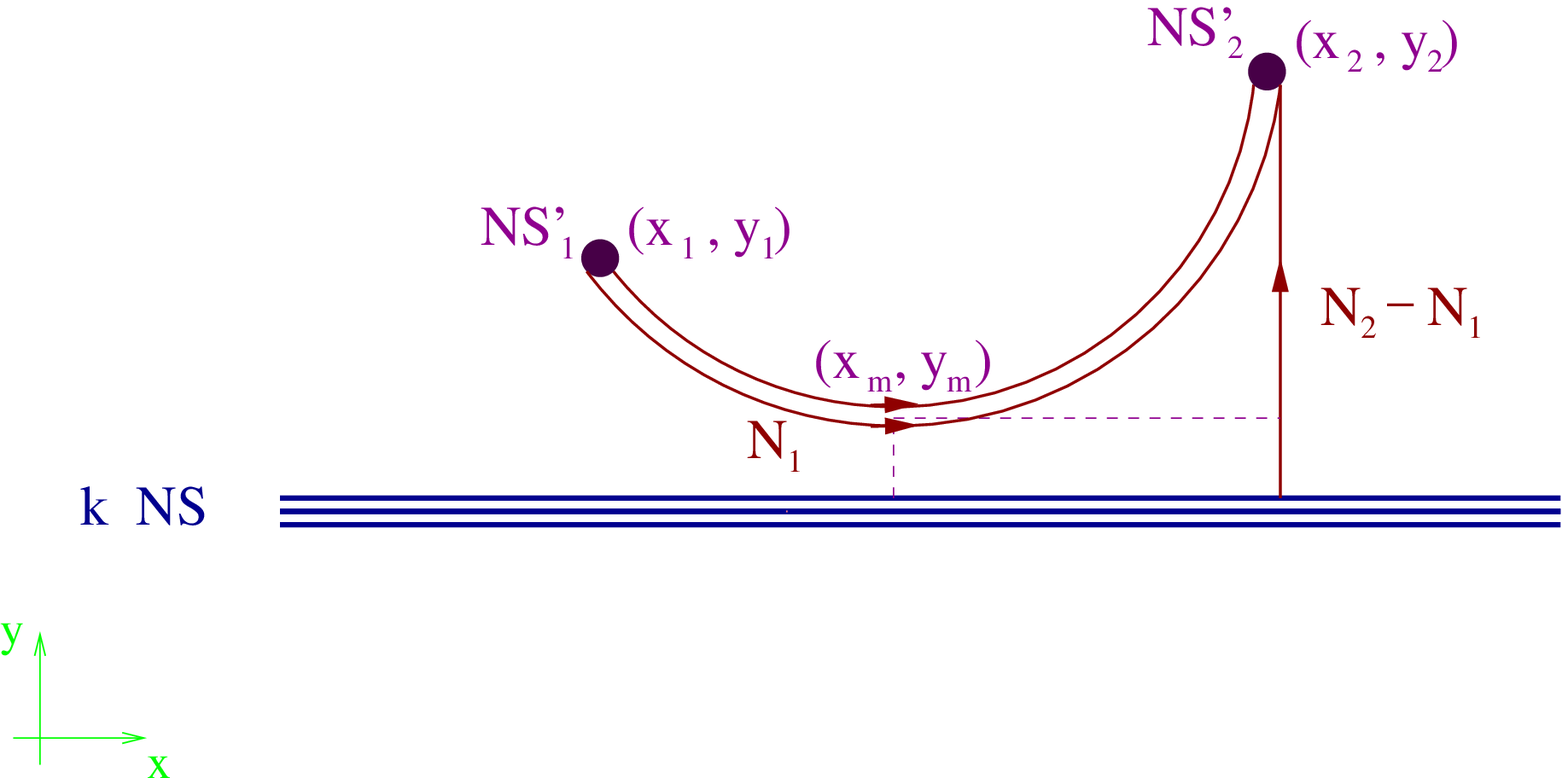}}

When $\Delta x$ exceeds a certain critical value (which can be
extracted from the formulae below) the branes take the qualitative
form in figure 6. In particular, the minimal value of $y$, $y_m$, is
attained at a point $x_m$ along the curve, $x_1<x_m<x_2$. In this
regime, which we will focus on here, the constant $C$ in \firstint\
is given by the value of the harmonic function \hcoin\ at $y_m$,
\eqn\expcc{C^2=H(y_m)~.}
Applying the first order differential equation \firstint\ to the two
intervals $x_1\le x\le x_m$ and $x_m\le x\le x_2$ leads to the
following two relations between the parameters in figure 6:
\eqn\intrel{\eqalign{\int_{y_m}^{y_1}{dy
H(y)\over\sqrt{H(y_m)-H(y)}}=&x_m-x_1~,\cr
\int_{y_m}^{y_2}{dy H(y)\over\sqrt{H(y_m)-H(y)}}=&x_2-x_m~.\cr}}
The integrals in \intrel\ can be performed exactly. One finds
\eqn\relfivesix{\eqalign{{y_m\over l}\sqrt{y_1^2-y_m^2}+l\theta_1=
x_m-x_1~,\cr
{y_m\over l}\sqrt{y_2^2-y_m^2}+l\theta_2= x_2-x_m~,\cr}}
where $\theta_i\in[0,{\pi\over2}]$, and
\eqn\defthetai{\cos\theta_i={y_m\over y_i}~.}
Adding the two equations \relfivesix\ and using \defthetai\ gives
the following relation between $\Delta x$ and $y_m$:
\eqn\finlfs{\Delta x=x_2-x_1={1\over 2l}\left(y_1^2\sin
2\theta_1+y_2^2\sin 2\theta_2\right)+l(\theta_1+\theta_2)~.}
The energy of the configuration of figure 6 is given by (minus) the
action \dbiradial,
\eqn\ecurved{E_{\rm curved}=N_1\tau_4\int dx \sqrt{ {1\over
H(y)}+(\partial_x y)^2} ~.}
Performing the integral in \ecurved, using \firstint, \expcc, one
finds
\eqn\lcurved{E_{\rm curved}=N_1\tau_4{\sqrt{H(y_m)}\over
2l}\left(y_1^2\sin2\theta_1+y_2^2\sin2\theta_2\right)~.}
To get a sense of the behavior described by equations \finlfs,
\lcurved\ we will next discuss in some detail the special case
$y_1=y_2=y$, and comment on the generalization to arbitrary
$y_2\geq y_1$.

\subsec{$y_1=y_2=y$}

In this case the two $NS'$-branes in figures 2, 6 are at the same
distance from the $NS$-branes. Equation \finlfs\ simplifies to
\eqn\ddxxyy{ \Delta x={y^2\over l}\sin2\theta+2l\theta~,}
with \defthetai\
\eqn\ddff{\cos\theta={y_m\over y}~.}
The angle $\theta$ provides a parametrization of $y_m$ in figure 6.
As it increases from $0$ to $\pi/2$, $y_m$ decreases from $y$ to
$0$. Thus, the relation \ddxxyy\ can be viewed as determining $y_m$
in terms of $\Delta x$. It turns out to be useful to discuss the
form of the solution separately for $y<l$ and $y>l$.

\ifig\loc{A plot of the function \ddxxyy\ for $l=\sqrt{2}y$.}
{\epsfxsize2.2in\epsfbox{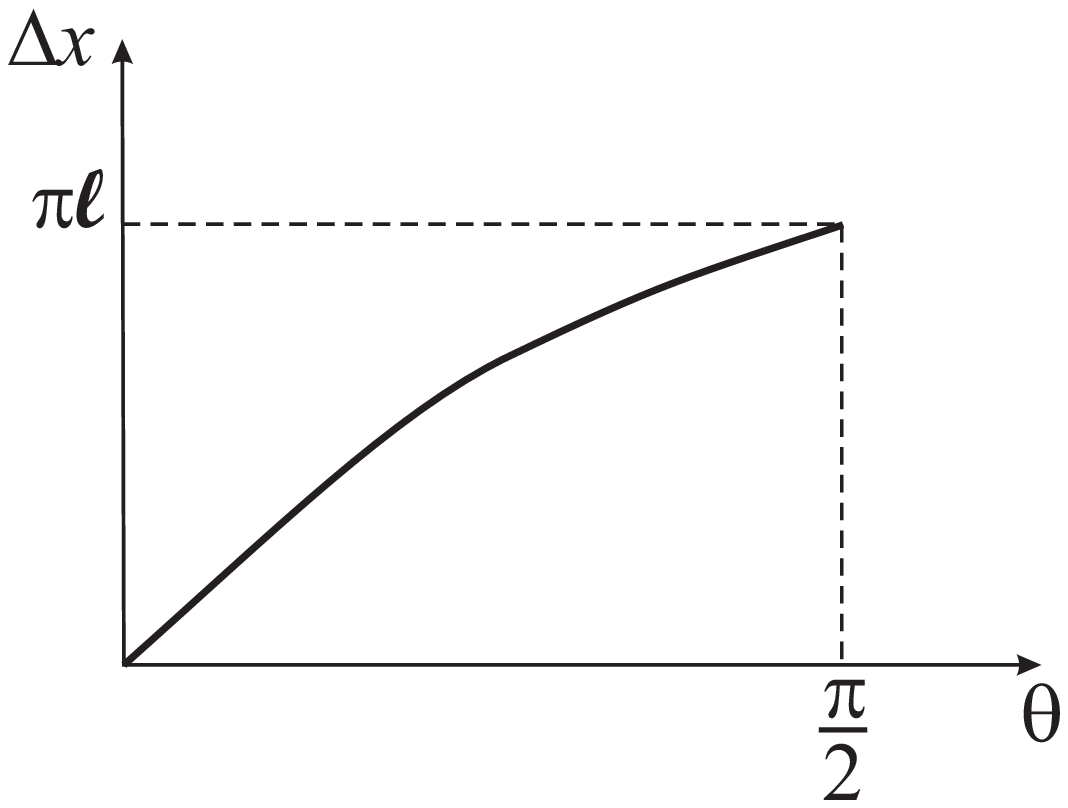}}

\subsubsec{ $y<l:$}

In this case the function \ddxxyy, which is plotted in figure 7 for
the special case $l=\sqrt{2}y$, is monotonically increasing. For
$\theta\to 0$ (\ie\ $y_m\to y$) it approaches $\Delta x=0$, while
for $\theta\to\pi/2$ ($y_m\to0$) it approaches $\Delta x=\pi l$. In
particular, in this case there are two distinct regimes:
\item{(1)} $0<\Delta x <\pi l$, where there are two solutions of the
equations of motion for the $D4$-branes: the straight brane solution
of figure 2 and the curved brane one of figure 6.
\item{(2)} $\Delta x\ge\pi l$, where the curved brane solution of
figure 6 does not exist, and the only allowed configuration is that
of figure 2.

\noindent In regime (1) one needs to determine the true ground
state, \ie\ to find which of the two allowed configurations has
lower energy. The energy of the $D4$-branes in figure 2 is given by
the flat space result
\eqn\straighte{E_{\rm straight}=N_1\tau_42y~.}
For the configuration of figure 6 we have
\eqn\curvede{E_{\rm curved}={N_1\tau_4\over l}\sqrt{H(y_m)}
y^2\sin2\theta=N_1\tau_42y\sqrt{H(y_m)}{y_m\over l}\sin\theta~.}
Dividing the two we find
\eqn\ratcur{\left(E_{\rm curved}\over E_{\rm
straight}\right)^2=\left[1+\left(y_m\over l\right)^2\right]
\left[1-\left(y_m\over y\right)^2\right]<1~.}
Thus, we see that for $y<l$, the configuration of figure 6 has lower
energy than the straight brane of figure 2, whenever it exists, \ie\
for all $\Delta x<\pi l$. This means that if we start with the
configuration of figure 2 with $\Delta x<\pi l$, and continuously deform
it\foot{In doing this, it is useful to think of $N_1$ of the $D4$-branes
in figure 2 as starting at the $NS_1'$-brane, going down to the $NS$-branes,
proceeding  along the $NS$-branes and then back up to the
$NS_2'$-brane. The segment that runs along the $NS$-branes is
massless to leading order in $g_s$.} in the direction of figure 6,
the energy decreases, until it reaches that of figure 6. If we
continue to deform it further, the energy increases again.

One can describe the situation qualitatively by the potential in
figure 8. The horizontal axis in that figure corresponds to one of
the many possible deformations of the shape of the $D4$-brane. For
example, one can think of it as labeling the smallest value of $y$
reached by the $D4$-branes. The configuration of figure 2
corresponds to the local maximum at the extreme left of figure 8,
while that of figure 6 corresponds to the global minimum of the
plot. As $\Delta x$ increases, this minimum moves to the left until,
at $\Delta x=\pi l$ it coincides with the local maximum at the
extreme left. Conversely, as $\Delta x$ decreases, the minimum moves
to the right, and as $\Delta x\to 0$ it approaches the boundary at
the extreme right of the plot, which corresponds to $\theta=0$ in
\ddxxyy\ or equivalently to $y_m=y$.

\ifig\loc{The qualitative behavior of the potential for the
$D4$-branes for $\Delta x<\pi l$.}
{\epsfxsize2.5in\epsfbox{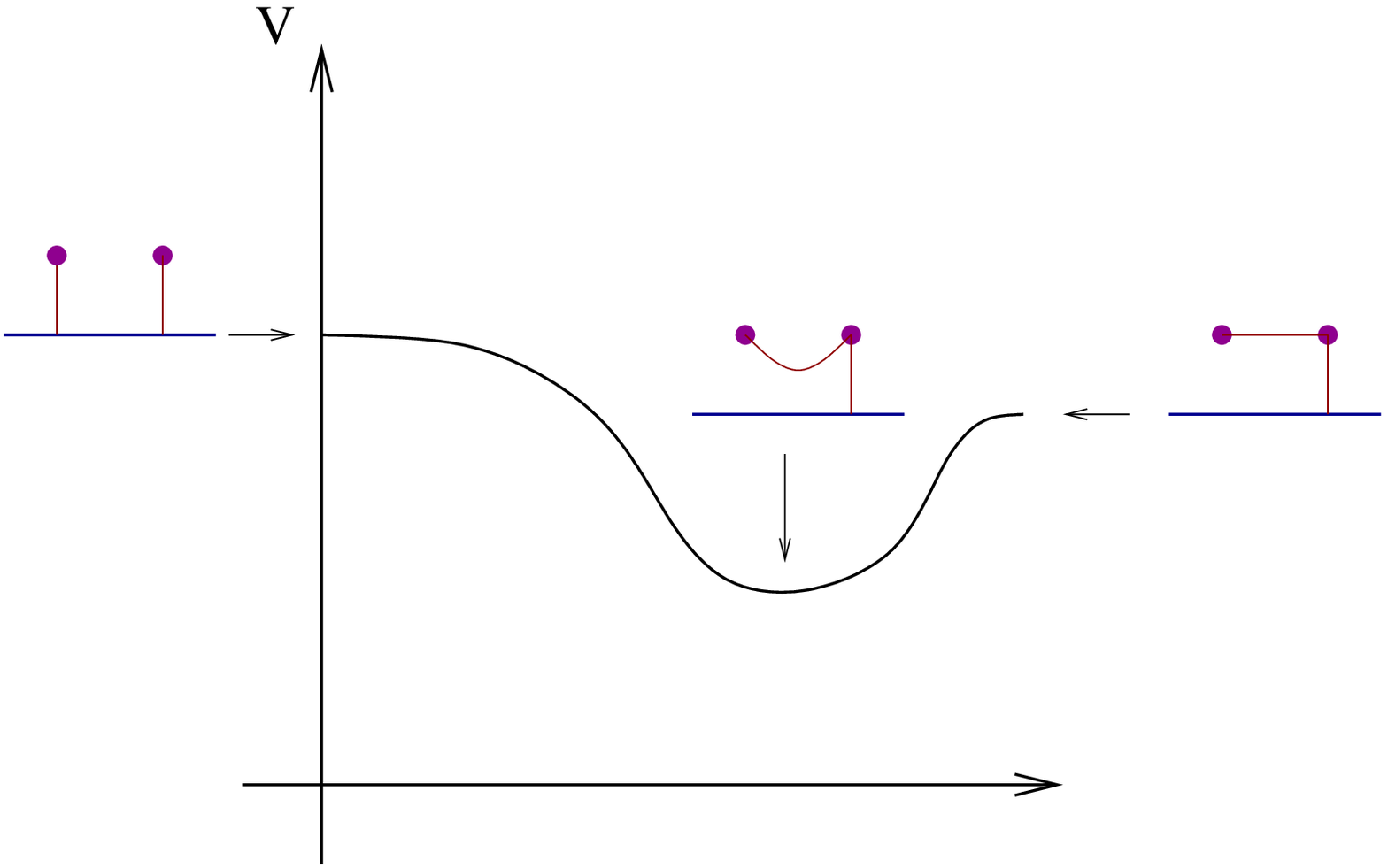}}

We see that for $\Delta x<\pi l$ the configuration of figure 2 is an
unstable equilibrium, \ie\ it contains a tachyonic mode. Above we
presented this mode as a geometric instability of the brane
configuration of figure 2 to deformations that take it towards that
of figure 6. One might feel uncomfortable with that presentation
since it involves the behavior of $D4$-branes arbitrarily close to the
$NS$-branes, deep in the strong coupling region near the fivebranes.

There is another, related, way of thinking about this instability,
which avoids this difficulty. It involves the dynamics of a
fundamental string stretched between the $D4$ and
$\overline{D4}$-branes in figure 2. Since this string is stretched
between a brane and an anti-brane, it satisfies the opposite GSO
projection from strings both of whose ends lie on $D$-branes with
the same charge. In particular, its lowest lying excitation is the
open string tachyon.

The mass squared of this tachyon receives a negative contribution
from the zero point energy of the string and a positive one from the
stretching of the string over a distance $\Delta x$. In the flat
space approximation of figure 2, if $\Delta x>\sqrt{2}\pi l_s$ the
tachyon is massive. We assumed above that $\Delta x<\pi l$, but
since $l$ \defll\ can in general be much larger than $l_s$, it
appears that there should be no instability associated with this
tachyon, at least for $\pi l>\Delta x>\sqrt{2}\pi l_s$.

In fact, the situation is more interesting. The $D4$ and
$\overline{D4}$-branes in figure 2 are stretched vertically from
$y=y_1=y_2$ all the way down to $y=0$. The effective mass of the
open string tachyon $T$, $m_T$, depends on position along the
branes. Deep inside the fivebrane throat, \ie\ for $y\ll l$, it can
be computed by using the near-horizon geometry of the fivebranes,
which is a linear dilaton space. The result is
\eqn\massnear{\left(2\alpha' m_T\right)^2=\left(\Delta
x\over\pi\right)^2-l^2~.}
The first term on the r.h.s. is due to stretching of the string in the
$x$ direction. The second provides a constant shift proportional to $k$. 
The derivation of equation \massnear\ is very similar to that of equation
(5.2) in \ItzhakiZR.

For comparison,\foot{The fact that the two masses coincide
for $k=2$ $NS$-branes seems to be related to observations in
\refs{\KutasovCT,\GiveonPR}.} the mass of the open string tachyon in flat
spacetime with constant dilaton is
\eqn\massflat{\left(2\alpha' m_T\right)^2=\left(\Delta
x\over\pi\right)^2-2l_s^2~.}
Equation \massnear\ implies that for $\Delta x<\pi l$, the open
string tachyon stretched between the branes and anti-branes in
figure 2 gives rise to a localized instability of the brane configuration
of figure 2. Its condensation leads to the configuration of figure 6.

For $\Delta x>\pi l$ the open string tachyon is massive everywhere,
and the configuration of figure 2 is locally stable. In fact, as we
have seen, it is the global ground state of the system in this
regime (at least classically).

Some comments are in order at this point:
\item{(1)} When the asymptotic string coupling is very small, there
is a wide range of distances in which the geometry \chs\ reduces to
a linear dilaton one, and the local string coupling is still small.
Thus, the instability of the configuration of figure 2 for $\Delta
x<\pi l$ can be seen at open string tree level.

\item{(2)} We have given above two descriptions of the instability of
the configuration of figure 2 to decay to that of figure 6. One is
purely geometric, associated with the discussion of figure 8. The
other involves the condensation of the tachyon stretched between
branes and anti-branes. In fact the two descriptions are known to be
related. In the near-horizon geometry of the fivebranes this is the
duality between the hairpin brane of \LukyanovNJ\ and the boundary
$N=2$ Liouville model (or boundary Sine-Liouville in the bosonic
case), which was discussed in the bosonic case in
\refs{\LukyanovNJ,\LukyanovBF} and in the fermionic case relevant
for our analysis in \KutasovRR. Our results suggest that this
duality can be extended to the full fivebrane geometry \chs.

\item{(3)} The discussion of the open string tachyon above is very
reminiscent of the study of non-supersymmetric deformations of the
CHS geometry in \ItzhakiZR. In fact, in the throat of the fivebranes
it is a direct open string analog of that system. As usual, in going
from closed to open strings the instability becomes easier to
analyze and the endpoint of tachyon condensation easier to identify.

To summarize, for $y<l$, as we vary $\Delta x$ the system undergoes
a phase transition. For $\Delta x<\pi l$, the configuration of figure 2
(which corresponds to the unbroken phase) is unstable, and the
stable one is that of figure 6. The order parameter can be taken to
be the expectation value of the tachyon $T$ stretched between the
branes and the anti-branes. It transforms in the bifundamental
representation of the gauge group on the branes \gggg, and is
non-zero in this regime. The gauge symmetry is broken,
\eqn\brokensym{U(N_1)\times U(N_2)\to U(N_1)_{\rm diag}\times
U(N_2-N_1)~.}
As $\Delta x$ increases, the order parameter $\langle T\rangle$
decreases; it goes to zero as $\Delta x\to\pi l$. For $\Delta
x>\pi l$ the order parameter vanishes and the theory is in the
unbroken phase, with the full gauge symmetry \gggg\ realized. Thus,
the system exhibits a second order phase transition at $\Delta x=\pi
l$. The behavior of the order parameter is schematically depicted in
figure 9.

\ifig\loc{The qualitative behavior of the order parameter for
$y<l$.} {\epsfxsize2.0in\epsfbox{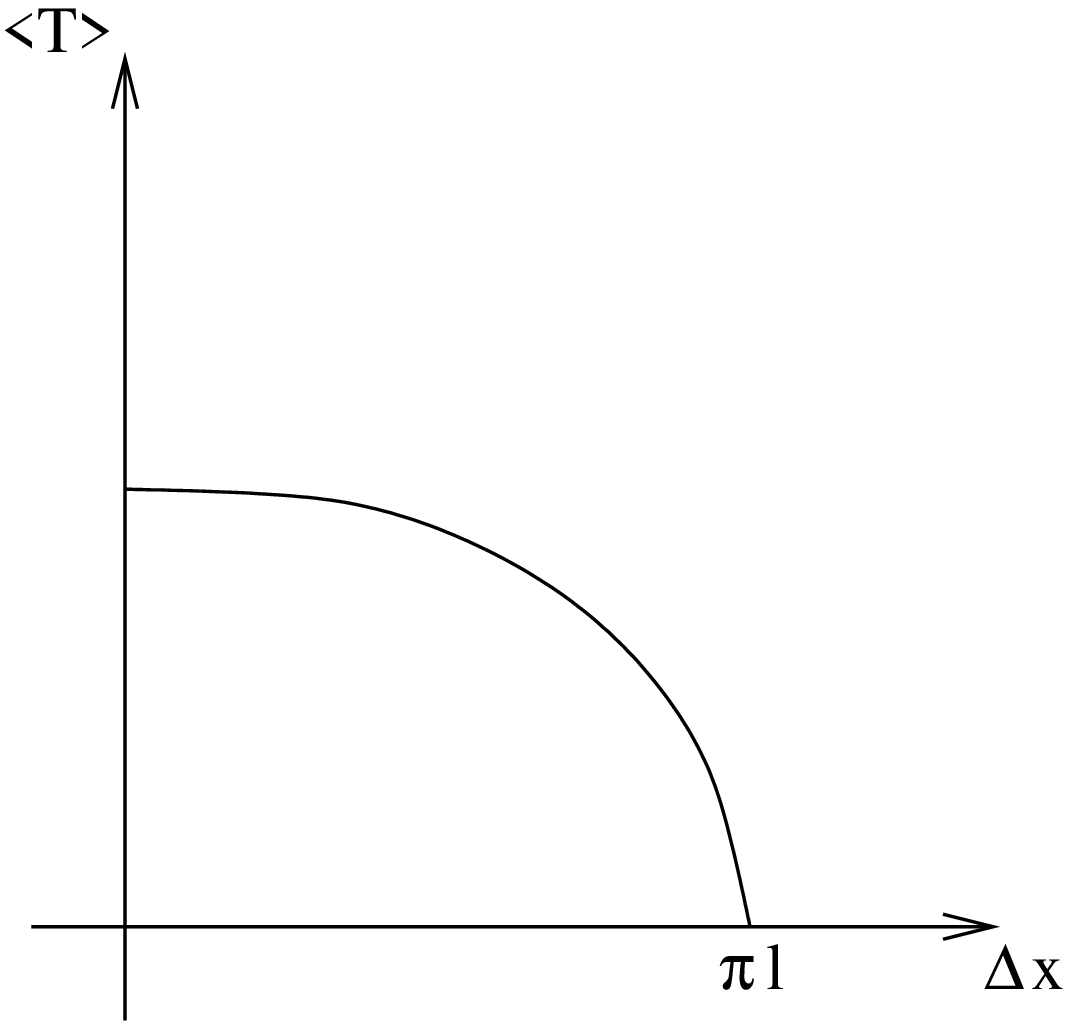}}

Geometrically, for small $\Delta x$ the ground state is the
configuration of figure 6 with $y_m$ slightly below $y$. As $\Delta
x$ increases, the curved $D4$-brane dips further and further towards
the fivebranes, and as $\Delta x\to\pi l$ it continuously approaches
the straight brane configuration of figure 2. For larger $\Delta x$
the configuration of figure 2 is the only possible one and is thus
the ground state of the system.

We stress again that the discussion above is classical. We will see
later that $g_s$ corrections may lead to additional vacua with unbroken
supersymmetry, to which all the states described here can decay. Nevertheless,
for small $g_s$ the picture we arrived at provides a very good approximation
to the physics of these states for a very long period of time.

\subsubsec{ $y>l:$}

Looking back at equation \ddxxyy, it is easy to see that in this
case $\Delta x$ is not a monotonic function of $\theta$ (see figure 10).
It has a maximum at $\theta=\theta_0$, with
\eqn\tthhee{\cos2\theta_0=-{l^2\over y^2}~.}
Note that ${\pi\over4}<\theta_0<{\pi\over2}$. The value of $y_m$
\ddff\ corresponding to \tthhee\ is
\eqn\valxsix{2y_m^2=y^2-l^2~.}
Plugging this into \ddxxyy, \ddff\ one can calculate the largest
$\Delta x$, $\Delta x_m$, for which a smooth solution exists. For
$y\gg l$ it is given by
\eqn\maxdelx{\Delta x_m \approx{y^2\over l}~.}
This should be compared to the case $y<l$, where we found that
$\Delta x_m=\pi l$. Like there, for $\Delta x>\Delta x_m$ there is a
unique solution to the DBI equations of motion -- the configuration
of figure 2. For $\Delta x<\Delta x_m$ the structure is more
intricate than before, and our next goal is to elucidate it.

\ifig\loc{A plot of the function \ddxxyy\ for $y=\sqrt{2}l$.}
{\epsfxsize2.2in\epsfbox{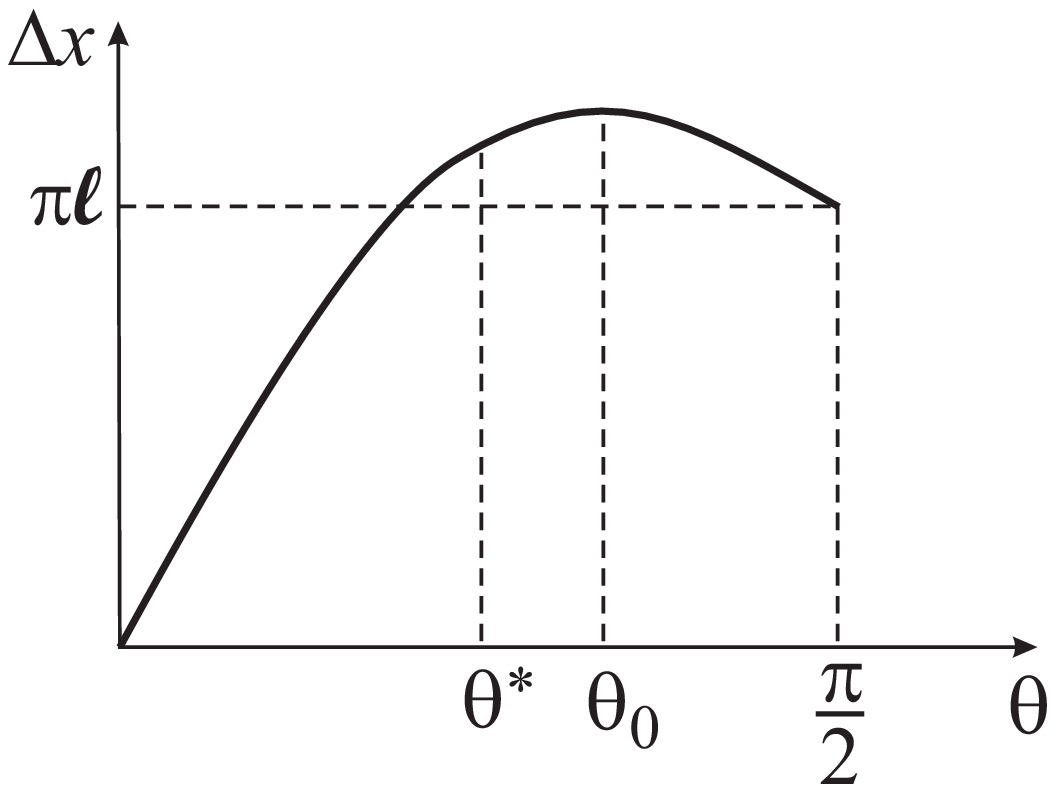}}

We start with the region $0<\Delta x<\pi l$. From figure 10 we see
that the DBI equations of motion have in this case a unique smooth
solution, and we expect its energy to be lower than that of the
straight brane solution of figure 2. The reason for that was
explained in subsection 3.2.1, where we pointed out that the open
string tachyon gives rise in this regime to an instability of the
configuration of figure 2, which is localized in the near-horizon
region of the fivebranes. Condensation of this tachyon should lead
to the configuration of figure 6, which therefore should have lower
energy.

To check this, we need to show that the ratio of energies in
equation \ratcur\ is smaller than one. One can rewrite this ratio as
\eqn\ratlarge{\left(E_{\rm curved}\over E_{\rm
straight}\right)^2=\left[1+\left(y\over
l\right)^2\cos^2\theta\right] \sin^2\theta~,}
and note that it has the following properties:
\item{(1)} It grows with $\theta$ in the range
$0<\theta<\theta_0$, where $\theta_0$  \tthhee\ is the value
corresponding to $\Delta x_m$ \maxdelx. For $\theta>\theta_0$ the
ratio \ratlarge\ starts decreasing.
\item{(2)} It is equal to one for $\theta=\theta^*$, with
\eqn\thetaone{\sin\theta^*={l\over y}~.}
By comparing \thetaone\ to \tthhee\ one finds that
$\theta^*<\theta_0$.
\item{(3)} At $\theta=\theta^*$, $\Delta x>\pi l$. Indeed, plugging
\thetaone\ into \ddxxyy\ we find that for $\theta=\theta^*$
\eqn\ccoott{{\Delta x\over l}=2(\cot\theta^*+\theta^*)~.}
The r.h.s. of \ccoott\ is a monotonically decreasing function of
$\theta^*$. It is equal to $\pi$ for $\theta^*=\pi/2$, and for all
other $\theta^*\in (0,{\pi\over2})$ it is larger. This establishes
that for $\Delta x<\pi l$ the smooth curved solution of the DBI
equations of motion has lower energy than the straight brane
configuration of figure 2, as expected.

\noindent
Having understood the region $\Delta x<\pi l$ we move on to $\Delta
x>\pi l$. According to figure 10, in this regime a second smooth
solution to the DBI equations of motion appears, at a larger value
of $\theta$, \ie\ a smaller value of $y_m$ (see \ddff). We will
denote the values of $\theta$ \ddff\ that correspond to the two
solutions of \ddxxyy\ by $\theta_S$ and $\theta_L$, respectively, where by
definition
\eqn\smth{\theta_S\le \theta_L~,}
\ie\ $\theta_S$ is the smaller of the two.

The fact that there are two solutions in this regime is simple to
understand from the preceding discussion. For $\Delta x>\pi l$, the
open string tachyon stretched between the $D4$ and
$\overline{D4}$-branes in figure 2 is massive (see \massnear). Thus,
this brane configuration is locally stable. However, if $\Delta x$
is only slightly larger than $\pi l$ it is not globally stable
since, as we have seen by analyzing the ratio \ratlarge, the energy
density of the configuration of figure 6 is smaller than that of
figure 2 for all $\theta<\theta^*$ \thetaone, which includes a
finite range of $\Delta x$'s larger than $\pi l$.

Therefore, if we start with the configuration of figure 2 and deform
it towards that of figure 6, we expect the energy to increase, reach
a maximum and then decrease to that of the global minimum, the
stable configuration of figure 6. The maximum of the energy between
the two minima corresponds to the second solution seen in figure 10.
We can describe the energetics by the qualitative plot in figure 11.

\ifig\loc{The qualitative behavior of the potential for the
$D4$-branes for $\Delta x>\pi l$, $\theta_S<\theta^*$.}
{\epsfxsize2.5in\epsfbox{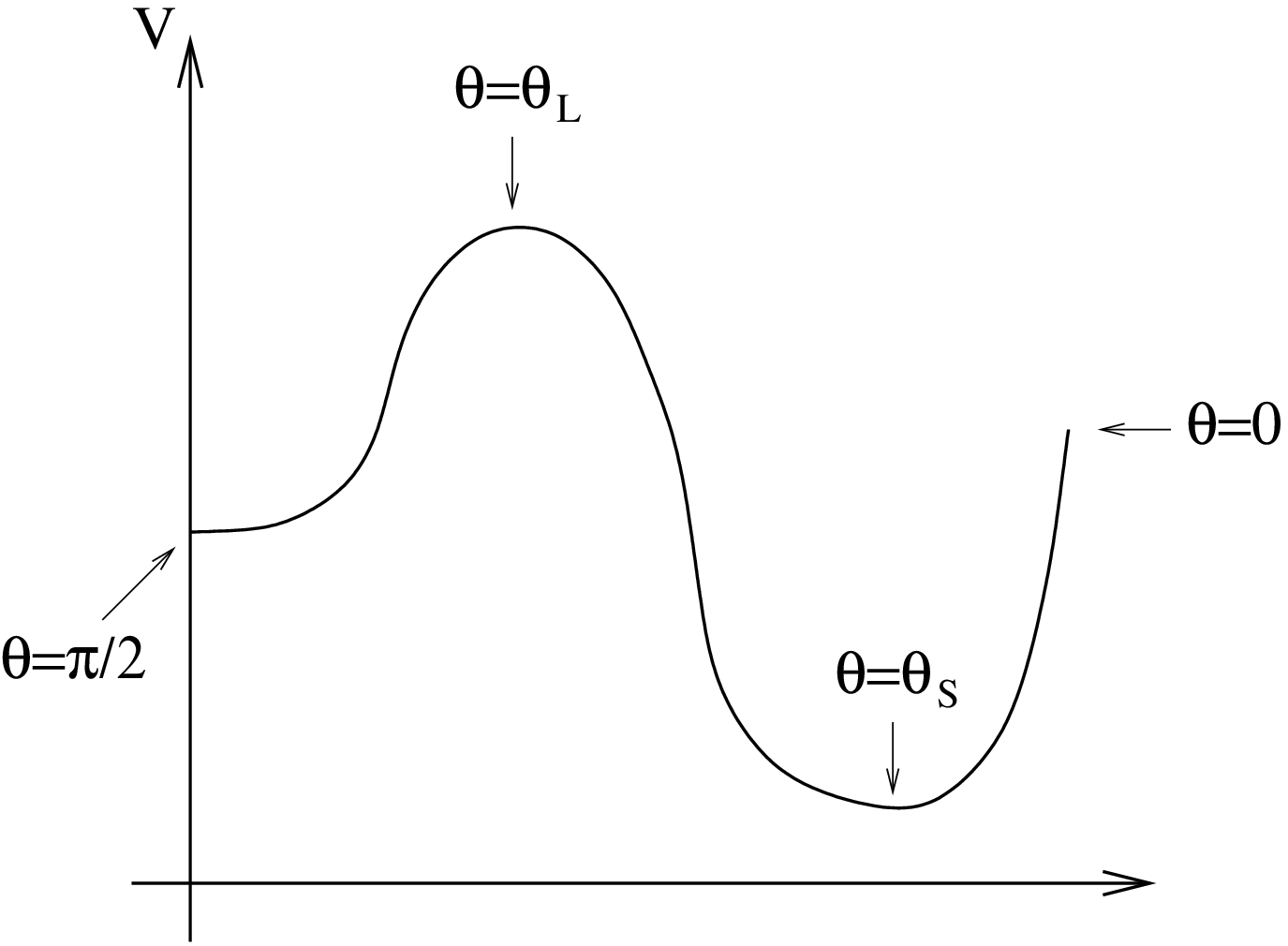}}

As is clear from this figure, the energy of the solution with
$\theta=\theta_L$ must be larger than that of the straight brane
configuration of figure 2 and than that of the solution with
$\theta=\theta_S$. Both of these expectations can be verified using
the formulae above.

The fact that the energy of the configuration with $\theta=\theta_L$
is larger than that of the one of figure 2 follows from our
discussion of the ratio \ratlarge\ above. We mentioned that this
ratio increases for $0<\theta<\theta_0$ and decreases for
$\theta_0<\theta<\pi/2$. Since $\theta_L>\theta_0$, the value of the
ratio \ratlarge\ at $\theta=\theta_L$ is larger than at
$\theta=\pi/2$. The latter is equal to one (essentially by
definition); therefore, the former is larger than one.

To prove that the energy for $\theta_L$ is larger than that for
$\theta_S$ one can proceed as follows. {}From \ratlarge\ we see
that we need to prove the inequality
\eqn\ssll{\sin^2\theta_S+\left(y\over
2l\right)^2\sin^22\theta_S\le \sin^2\theta_L+\left(y\over
2l\right)^2\sin^22\theta_L~.}
This is equivalent to
\eqn\eqssll{\cos 2\theta_S+\left(y\over
2l\right)^2\cos4\theta_S\ge \cos 2\theta_L+\left(y\over
2l\right)^2\cos4\theta_L~.}
Rearranging this inequality and using a trigonometric identity
for the difference of cosines gives
\eqn\ineqcos{\sin(\theta_S+\theta_L)\sin(\theta_L-\theta_S)\ge
-\left(y\over
2l\right)^2\sin[2(\theta_S+\theta_L)]\sin[2(\theta_L-\theta_S)]~.}
Using the doubling formula for sine and the fact that
$\theta_S+\theta_L$ is between $0$ and $\pi$ so
$\sin(\theta_L\pm\theta_S)$ is positive, gives
\eqn\ppww{1\ge -\left(y\over
l\right)^2\cos(\theta_S+\theta_L)\cos(\theta_L-\theta_S)~.}
We also need to use the fact that $\theta_S$ and $\theta_L$
correspond to the same $\Delta x$ \ddxxyy. This means that
\eqn\samelf{\left(y\over l\right)^2\cos(\theta_S+\theta_L)
{\sin(\theta_L-\theta_S)\over  \theta_L-\theta_S}=-1~.}
Plugging this in \ppww\ we conclude that we need to prove that
\eqn\rrtt{(\theta_L-\theta_S)\cot(\theta_L-\theta_S)\le 1~.}
Equivalently, we need to show that
\eqn\fftt{f(\theta)=\tan\theta-\theta}
is positive for $\theta=\theta_L-\theta_S\in(0,{\pi\over2})$. This
is indeed the case, since $f$ vanishes at $\theta=0$ and its
derivative is positive for all ${\pi\over2}>\theta>0$.

The above checks substantiate the picture suggested by figure 11 for
$\Delta x>\pi l$ but sufficiently small such that
$\theta_S<\theta^*$ \thetaone. As we increase $\Delta x$ beyond that
point, we get to a regime where, as we have seen, the energy of the
straight brane configuration of figure 2 is lower than that of the
smooth curved solution to the DBI equations of motion. In this
regime, the potential has the qualitative structure depicted in
figure 12.

\ifig\loc{The qualitative behavior of the potential for the
$D4$-branes for $\Delta x>\pi l$, $\theta_0>\theta_S>\theta^*$.}
{\epsfxsize2.5in\epsfbox{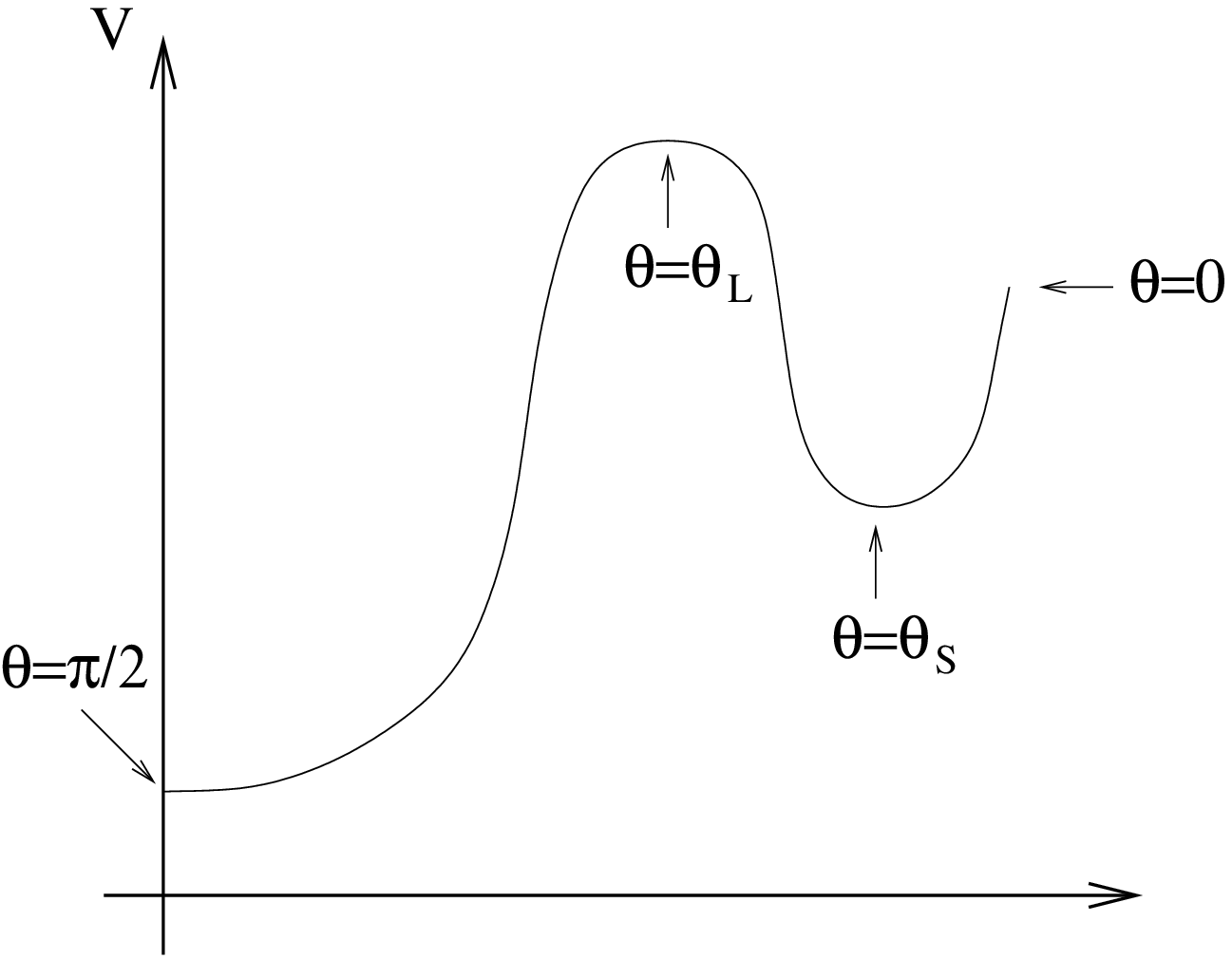}}

Thus,  for $\theta_S<\theta^*$ (or small $\Delta x$) the classical ground
state of the system corresponds to a brane configuration of the type
of figure 6. The gauge symmetry is broken as in \brokensym, and the
corresponding order parameter, the expectation value of the
bifundamental tachyon, is non-zero. On the other hand, for
$\theta_S>\theta^*$ (sufficiently large $\Delta x$), the ground
state is the configuration of figure 2, the order parameter
vanishes, and the gauge symmetry is unbroken.

The transition between the two regimes at $\theta=\theta^*$ is a
first order phase transition. The order parameter jumps
discontinuously from a non-zero value to zero as we pass that point
(see figure 13).

\ifig\loc{The qualitative behavior of the order parameter for
$y>l$.} {\epsfxsize2.0in\epsfbox{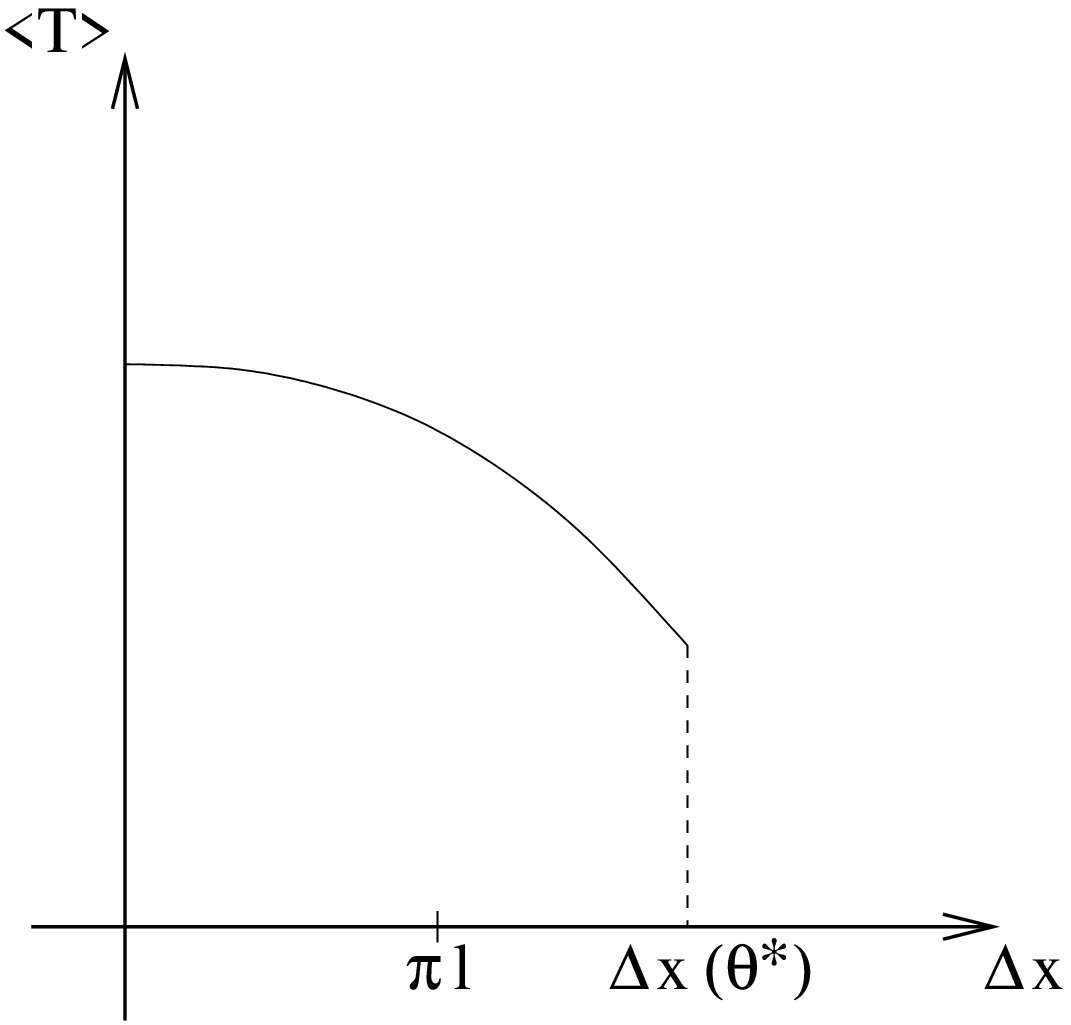}}

For $\theta_0>\theta_S>\theta^*$, the broken vacuum remains a local minimum
of the potential and is a meta-stable state at weak string coupling.
As we increase $\theta_S$ (or $\Delta x$), the local minimum at
$\theta_S$ and local maximum at $\theta_L$ approach each other, and the
meta-stable state with broken symmetry becomes less stable.
Eventually, at a value of $\Delta x$ corresponding to
$\theta_S=\theta_L=\theta_0$, the meta-stable minimum and local
maximum meet and for larger values of $\Delta x$ the equations of
motion no longer have a smooth curved solution of the kind drawn in
figure 6.

It is instructive to compare the preceding discussion to the one of
section 2.2, where we effectively took $l\to 0$. There, we found a
transition at \posneg\ $\Delta x=2y$. Here, the transition was found
to occur at $\theta=\theta^*$ \thetaone. To compare to section 2 we
need to take $y\gg l$, and plugging this into \thetaone, \ddxxyy\ we
find the same answer as there. An important qualitative difference
with respect to the analysis of section 2 is that there the broken
vacuum of figure 3 appeared to be a meta-stable state for arbitrarily
large $\Delta x$. The more accurate analysis of this section
revealed that this is true for $\Delta x<y^2/l$ \maxdelx, but beyond
that point the meta-stable broken vacuum state ceases to exist. This
is consistent with the discussion of section 2, since the upper
bound \maxdelx\ goes to infinity as $l\to 0$.

\subsec{General $y_1$, $y_2$}

In this subsection we comment on the generalization of the results
of the previous subsection to arbitrary $y_1\le y_2$, for which the
relation between $\Delta x$ and $y_m$ in figure 6 is given by eq.
\finlfs. In subsection 3.2 we found it convenient to parametrize
$y_m$ via the angle $\theta$ \ddff. The natural analog of $\theta$
in the general case is $\theta_1$ in
\defthetai. As this angle varies between $0$ and $\pi/2$, $y_m$
varies between $y_1$ and $0$, respectively. The latter case
corresponds to the geometry of figure 2. The angle $\theta_2$,
which appears in equation \finlfs, can be expressed in terms of
$\theta_1$ using
\defthetai\ as
\eqn\thetatwo{\cos\theta_2={y_1\over y_2}\cos\theta_1~.}
The resulting function $\Delta x(\theta_1)$ \finlfs\ is a natural
generalization of \ddxxyy\ to arbitrary $y_1\le y_2$. Its values at
the boundaries $\theta_1=0,\pi/2$ are the following. For
$\theta_1=\theta_2=\pi/2$, one finds $\Delta x=\pi l$ (this is the
configuration of figure 2, which of course exists for all $\Delta x$,
but is only obtained as a limit of a
smooth solution to the DBI equations of motion as $\Delta x\to\pi l$).
The value of $\Delta x$ for $\theta_1=0$, which corresponds to $y_m=y_1$,
depends on $y_2$, and grows linearly with $y_2$ for large values of the
latter. Large $y_2$ and small $\Delta x$ is an example of a regime
(mentioned in section 3.1) in which $y(x)$ is monotonic. We will not
study this regime here.

As in subsection 3.2, it is interesting to determine whether the
function $\Delta x(\theta_1)$ is monotonic in its regime of
validity. A short calculation shows that it is monotonically
increasing for
\eqn\monot{y_1y_2<l^2~.}
For $y_1y_2>l^2$ it has a maximum at
\eqn\nonmonot{\cos(\theta_1+\theta_2)=-{l^2\over y_1 y_2}~.}
These results generalize those found in the previous subsection for the
case $y_1=y_2=y$. In particular, \nonmonot\ is the analog of \tthhee\ for
general $y_1,y_2$.

As before, it is useful to discuss separately the regime \monot\ and its
complement. Much of the analysis is similar to that of subsection 3.2, so
we will be brief.

\subsubsec{ $y_1y_2<l^2$}

For $\Delta x>\pi l$ there is again a unique solution to the
equations of motion, corresponding to the straight branes of
figure 2. For $\Delta x<\pi l$ there are two solutions, those of
figures 2 and 6, and we need to compare their energies. For
reasons that were explained in the previous subsection, we expect
the smooth, curved solution of figure 6 to be the true classical
ground state in this case. To show this, we must prove that (see
\lcurved)
\eqn\engeny{{1\over
2l}\sqrt{H(y_m)}\left(y_1^2\sin2\theta_1+y_2^2\sin2\theta_2\right)<y_1+y_2~.}
Using \defthetai, this can be rewritten as
\eqn\newcond{F(y_m)<y_1+y_2~,}
where
\eqn\deffm{F(y_m)=\sqrt{1+{y_m^2\over
l^2}}\left(\sqrt{y_1^2-y_m^2}+ \sqrt{y_2^2-y_m^2}\right)~.}
To see that the inequality \newcond\ is indeed correct, one notes
that it becomes an equality for $y_m=0$ (by construction), and for
any larger $y_m$ the function \deffm\ is smaller since the
derivative ${dF\over dy_m}$ is negative. Thus, we conclude that,
as expected, for $\Delta x<\pi l$ the configuration of figure 2 is
unstable to the condensation of the tachyon discussed in the
previous subsection, whose mass is given by \massnear, and the
stable configuration is that of figure 6.

As $\Delta x\to \pi l$, the configuration of figure 6 approaches
that of figure 2, and for larger $\Delta x$ there is a unique
vacuum. The system undergoes a second order phase transition at
$\Delta x=\pi l$, as in figure 9.

\subsubsec{ $y_1y_2>l^2$}

We expect a similar picture to that of subsection 3.2. For $\Delta
x<\pi l$ there should be a unique smooth curved solution to the DBI
equations of motion. For $\Delta x$ slightly above $\pi l$ a second
solution with $y_m\ll y_1, y_2$ should appear. This solution should
be a local maximum of the energy, separating the global minimum of
figure 6 from the local one (figure 2).

The appearance of a solution with small $y_m$ can be verified
directly as follows. Equation \defthetai\ implies that
\eqn\leadord{\theta_i={\pi\over2}-{y_m\over y_i}+O(y_m^2)~.}
Plugging this into \finlfs\ one finds
\eqn\smallymm{\Delta x=\pi l+y_m\left({y_1\over l}+{y_2\over
l}-{l\over y_1}-{l\over y_2}\right)+O(y_m^2)~.}
The expression in brackets in \smallymm\ is positive for
$y_1y_2>l^2$, so when $\Delta x$ is slightly larger than $\pi
l$, the solution for $y_m$ is small, as expected. In order to
show that this solution is a local maximum of
the energy, one has to check that the function
$F(y_m)$ \deffm\ satisfies $F(y_m)>y_1+y_2$. It is easy to check
that this is indeed the case to leading order in $y_m$.

The general picture in this case is expected to be very similar to
that of subsection 3.2.2. For $\Delta x$ slightly above $\pi l$, the
straight brane configuration of figure 2 is a local minimum of the
energy, while the true ground state is a curved configuration as in
figure 6. As $\Delta x$ increases, the relative energies of the two
minima change, until when the inequality \newcond\ becomes an
equality, they flip and the straight brane configuration becomes the
ground state. The configuration of figure 6 remains a local minimum
until a much larger value of $\Delta x$, above which it ceases to
exist. We will leave a more detailed investigation to future work.

\subsec{ Finite $k$ corrections}

The analysis of this section so far involved two types of approximations.
First, we used the supergravity result for the fivebrane geometry, \chs.
In principle there can be perturbative and non-perturbative $\alpha'$
corrections to the background. Second, we used the DBI approximation
to describe the D-branes, and again in general one expects corrections
to this description. For large $k$, both of these approximations are
justified, but one can ask what happens for finite $k$. In this subsection
we briefly comment on this issue.

Some features of our description are known not to receive $\alpha'$ corrections.
In particular, the near-horizon geometry of $k$ fivebranes, which is obtained
from \chs\ by omitting the constant term in the harmonic function $H$, \hcoin,
is known to be valid for all $k\ge 2$ \refs{\CallanAT\AharonyUB-\AharonyXN}. A
corollary of this is that our assertion that the brane configuration of figure 2
develops a localized tachyon for $\Delta x<\pi l$, with $l$ given by \defll, is
exact as well. Indeed, regardless of the values of $y_1$ and $y_2$ in figure 2,
the $D4$ and $\overline{D4}$-branes always stretch all the way down the semi-infinite
fivebrane throat. The linear dilaton description of this throat \CallanAT\ leads
to the mass formula \massnear\ which is thus exact.

In the near-horizon, linear dilaton geometry, the phase structure
of the full string theory must agree with our DBI analysis of the
case $y<l$ in subsection 3.2.1. To recapitulate, the system
undergoes a second order phase transition at $\Delta x=\pi l$. In
the broken phase the shape of the $D4$-branes is a piece of the
hairpin brane\foot{Or, more precisely, of its $N=2$ superconformal
generalization discussed in
\refs{\KutasovDJ,\NakayamaYX,\KutasovRR}.} of
\refs{\LukyanovNJ,\LukyanovBF}, which exists only for $\Delta
x<\pi l$. As $\Delta x$ approaches $\pi l$ from below, the bottom
of the hairpin approaches the fivebranes and the solution smoothly
connects to that of figure 2. The order parameter behaves
qualitatively as in figure 9. The fact that the hairpin brane is
an exact solution of the classical open string equations of motion
implies that the above statements are valid for all $k\ge 2$.

Another part of our analysis that is valid for all $k$, including
$k=1$, involves the behavior of the solution of figure 6 for
$y_i\gg l$ (in the full, asymptotically flat geometry \chs). In
this regime the solution of figure 6 is located entirely in the
large $y$ region. Indeed, even for the largest separation $\Delta
x$ for which it exists, corresponding to $\theta=\theta_0$
\tthhee, the minimal value of $y$ along the $D4$-brane, $y_m$
\valxsix, is in this case large. Therefore, the DBI approximation
is valid not necessarily because $k$ is large but because the
$D$-brane sits in a region of small curvature at large $y$.

Thus, the fact that for large $y_i$ the system undergoes a strongly first order
phase transition at a value of $\Delta x$ approximately given by \ddxxyy, \thetaone,
is a reliable outcome of our analysis. As $y_i$ decrease towards $l$ the phase
transition becomes less strongly first order. An interesting question is whether
there is a critical value of $y$ below which the transition becomes second order,
as was found in our DBI analysis, or whether the transition remains first order for 
arbitrarily small $y$  due to $1/k$ corrections, only becoming second order as $y\to 0$.

We believe that for small $y$ the phase transition must be second order, as in the DBI
analysis. Indeed, consider the brane configuration of figures 2, 6 for $y_1=y_2=y\ll l$.
Since the configuration of figure 2 is locally stable only for $\Delta x\ge \pi l$, any
first order transition would have to occur at a value of $\Delta x$ larger than $\pi l$.
This means that there must be a smooth solution of the sort depicted in figure 6 for
$\Delta x$ slightly larger than $\pi l$. At the same time, this solution should extend into
the region in which the fivebrane geometry is well approximated by the linear dilaton one.
Hence it has to look locally like the hairpin brane, which has by construction $\Delta x<\pi l$.

To summarize, it appears that the DBI analysis of this section
captures the correct phase structure of the full classical string
theory in our background, and describes correctly many
quantitative features as well. Some parameters, such as the value
of $y$ at which the phase transition goes from being first order
to second order, can receive $\alpha'$ corrections. Others, such
as the value of $\Delta x$ at which the configuration of figure 2
ceases to exist, are given precisely by the DBI analysis. It is
also important to remember that the discussion here is entirely
classical. We will see below that quantum corrections give rise to
qualitatively new features.

\newsec{Gauge theory analysis at small $\Delta x$}

In the previous section we studied the dynamics of the brane
configuration of figure 2 in classical string theory. It is
interesting to analyze the quantum $(g_s)$ corrections to the
resulting picture. For small $\Delta x$ this can be done by
using the low energy field theory on the branes, which is well
understood. In this section we will review the dynamics
of this field theory. In the next section we will suggest the
generalization for larger $\Delta x$.

We start with the brane configuration of figure 5. For simplicity,
we restrict the discussion to the case where the number of $NS$-branes
is $k=1$.\foot{It is possible to generalize it to larger $k$ by using
the relevant gauge theory for that case \GiveonSR. This is necessary
for making contact with the analysis of section 3, much of which assumes that
$k>1$.} The low energy dynamics of this brane configuration is described
by an $N=1$ supersymmetric gauge theory with gauge group
\eqn\gmag{G_{\rm mag}=U(N_2-N_1)\times U(N_2)\equiv U(\widetilde N_1)\times U(N_2)~.}
The gauge couplings of the two factors are given by
\eqn\coupmag{\tilde\alpha_1={g_sl_s\over y_1}~,\qquad\qquad
\tilde\alpha_2={g_sl_s\over y_2-y_1}~.}
In particular, in the regime described in the previous sections, small $g_s$ and
fixed $y_i/l_s$, the classical gauge theory is weakly coupled. By tuning $y_1$ and $y_2$
one can arrange for one of the gauge couplings to be much larger than the other, so
that only one of the factors in the gauge group \gmag\ is important.

In addition to the gauge multiplets, the gauge theory contains the following chiral
superfields: an adjoint of $U(N_2)$, $\Phi_2$ (which is proportional to $M$ \cubsup),
and bifundamentals $q$, $\tilde q$ in the $(\widetilde N_1, \overline N_2)$ and
$(\overline{\widetilde N_1}, N_2)$, respectively. The superpotential is given by
\eqn\www{W=h\M \tilde q q~,}
where
\eqn\hhh{h^2=\tilde\a_2~.}
The deformation of figure 5 that appears in figures 2, 3, corresponding to a relative
displacement of the $NS'$-branes in the $x$ direction, is described in the above gauge
theory by an addition of a linear term to the superpotential \www. Following
\refs{\IntriligatorDD,\BenaRG} we parametrize it as follows:
\eqn\wto{W=h\M \tq q -h\mu^2\Tr\,\M~.}
The mass parameter $\mu$ is given in terms of $g_s,l_s$ and $\dx$
by (see eq. (2.11) in \BenaRG)
\eqn\mmxgl{\mu^2=-{\dx\over 2\pi g_sl_s^3}~.}
The above gauge theory is valid at energy scales well below a scale $E_c$
which can be taken to be the lowest of $m_s$ and the Kaluza-Klein scales $1/y_i$.

Classically, the F-term of $\Phi_2$ in \wto\ leads to
supersymmetry breaking via a mechanism that generalizes the
O'Raifeartaigh model. Some of the components of $\Phi_2$ get a
mass in the process; some others give rise to pseudo-moduli, which
have a classically flat potential. The supersymmetry breaking
ground state is described in terms of branes by the configuration
of figure 3 \refs{\OoguriBG\FrancoHT-\BenaRG}. The pseudo-moduli
correspond to translational modes of the $D4$-branes stretched
between the $NS'$-branes.

Quantum mechanically, two things happen. The potential of the
pseudo-moduli ceases to be flat near the origin, and
supersymmetric vacua appear at finite $\Phi_2$. When the coupling
$\tilde\alpha_2$ in \coupmag\ is small, the analysis of these
effects is identical to that of \IntriligatorDD. In particular,
the mass of the pseudo-moduli is of order\foot{Equations \coupmag,
\hhh\ and \mmxgl\ imply that this mass is proportional to
$\sqrt{g_s}=g_{\rm open}$. This is consistent with the fact that
it comes from one loop open string effects (\ie\ the annulus).}
$h^2|\mu|$. The gauging of $U(N_2)$ leads to corrections to that
analysis, but these are small if the dynamically generated scale,
\eqn\lbb{\lb=E_c\exp\left[-{8\pi^2(y_2-y_1)\over(2N_2-\widetilde
N_1)g_sl_s}\right]~,}
is much smaller than the mass $h^2|\mu|$. This can be achieved by increasing $y_2$, thereby
decreasing the gauge coupling $\tilde\alpha_2$ \coupmag.

The supersymmetric vacua occur at
\eqn\hm{\langle
h\M\rangle\simeq\left(\mu^{2\widetilde N_1}\la^{N_1-2\widetilde N_1}\right)^{1\over N_1}~.}
We are interested, following \IntriligatorDD, in the case $N_2>3\widetilde N_1$, where
the $U(\widetilde N_1)$ factor in \gmag\ is not asymptotically free. It
becomes strongly coupled at the scale
\eqn\laa{\la=E_c\exp\left[8\pi^2y_1\over(N_2-3\widetilde N_1)g_sl_s\right]~.}
In the regime of interest for our analysis this scale is much
higher than $E_c$, so the gauge theory description breaks down when it is
still weakly coupled. In particular, for the analysis of \IntriligatorDD\ to
be valid, one must have
\eqn\hmlll{\langle h\M\rangle\ll E_c~.}
Plugging in the values of the different parameters one finds that \hmlll\ implies
that the gauge theory analysis is only valid in the regime where $\Delta x$
is smaller than $\exp(-C/g_s)$ for some positive constant $C$.

We see that the DBI regime of section 3 is well outside the regime of validity of the
gauge theory. The relevant description in this regime is in terms of brane dynamics in
string theory, which provides a different UV completion of the non-asymptotically free
magnetic gauge theory \gmag\ than that of \IntriligatorDD. There, the UV theory is the
electric theory \gggg\ described in section 2. Here, it the open+closed string theory
in the background of figure 5 (and its deformations). However, in the regime of validity
of the magnetic gauge theory \hmlll, the vacuum structure is insensitive to the UV
completion and the above analysis is valid.

\newsec{Quantum effects at large $\Delta x$}

In the previous section we analyzed the low energy dynamics of the full quantum theory
corresponding to the brane configuration of figure 3 using the gauge theory description,
which is valid for small $\Delta x$ (or small $m$ \formm). In this section we will discuss
quantum effects in the regime studied classically in section 3, where $\Delta x$ is of
order $l_s$ or larger, and $g_s$ is very small. As we will see, a full analysis of this
problem requires more work, but we will suggest a picture that incorporates the
results of sections 3, 4 and interpolates between them.

An important phenomenon that is absent classically but needs to be
taken into account in the quantum analysis is bending of the
$NS5$-branes \WittenSC. Consider, for example, the $NS'_1$-brane
in figure 2. The $N_1$ $\overline{D4}$-branes ending on it are
seen in the fivebrane worldvolume theory as charged particles in
the two remaining dimensions along the fivebrane,
\eqn\remdim{w=x^8+ix^9~.}
Classically, the $NS'_1$-brane is located at a particular value of $y$, $y=y_1$. For finite
$g_s$ its location in $y$ becomes a function of the distance from the fourbranes, $|w|$
\WittenSC. Asymptotically, for large $|w|$, it behaves like
\eqn\largewbend{y\simeq N_1g_sl_s\ln{|w|\over l_s}~.}
There are also some other fields on the fivebrane that behave non-trivially due to the presence
of the $D4$-branes, but we will not discuss them in detail.

The bending \largewbend\ is in the $y$ direction since this direction lies along the $D4$-branes
and transverse to the $NS'$-brane. Thus, it depends on the orientation of the $D4$-branes
(as well as on their number). This means that in going from the configuration of figure 2 to
that of figure 3, which as we argued above is sometimes energetically favorable, the asymptotic
shape of the $NS'$-branes at large $|w|$ changes by an infinite amount. At first sight this
seems to suggest that such transitions are dynamically impossible, and one should take the
shape of the fivebranes at infinity as given when studying these systems \BenaRG.

Our view is that processes in which the asymptotic bending of the fivebranes changes {\it are}
dynamically allowed. Consider for example the case (analyzed classically in section 3) where
the two $NS'$-branes are located at $y_1=y_2=y\gg l$, and the distance between them, $\Delta x$,
is in the range $\sqrt{2}\pi l_s<\Delta x<\pi l$. In this case, we saw that the open string tachyon
leads to a localized instability of the $D-\bar D$ system in the near-horizon region of the
$NS$-branes. Indeed, while its mass squared in the flat space far from the fivebranes, \massflat,
is positive, the near-horizon mass squared \massnear\ is negative.

If the $D4$-branes and antibranes were parallel and extended all
the way to infinity in the $y$ direction, the above localized
tachyon would condense, and as a result the branes and antibranes
would connect and move off to large $y$. In the system of figure
2, the initial stage of the process of tachyon condensation has to
be the same, since the dynamics near the $NS$-branes has no way of
knowing that at some large value of $y$ the $D4$-branes and
antibranes are attached to other branes. Thus, one expects the
tachyon localized near the $NS$-branes to start condensing,
leading to a reconnection of the D-branes deep in the throat of
the fivebranes. The part of the D-branes where they attach to the
$NS'$-branes is initially not influenced by the reconnection.

At some point in the process of tachyon condensation, when the D-branes already have a shape similar
to that of figure 6, this part of the D-branes is sufficiently deformed that the $NS'$-branes backreact.
They develop a kink which interpolates between the behavior \largewbend\ and the one appropriate for
the smooth connected solution of figure 6. This kink moves off to infinity such that at late times the
full configuration looks like that of figure 6, with the bending appropriate for the D-brane shape. Of
course, this process takes an infinite amount of time, but for an observer localized near the intersection
this is irrelevant, since the kink leaves the vicinity of the intersection very rapidly.

Some comments about the preceding discussion are in order:

\item{(1)} In section 3 we found that in some cases the instability of a brane configuration
such as that of figure 2 is non-perturbative and requires tunneling through a barrier. The
discussion above should apply to these as well. Consider for example the case where
the $NS'$-branes are at a $y$ slightly larger than $l$, and the distance between the $D4$ and
$\overline{D4}$-branes, $\Delta x$, is slightly above $\pi l$. We saw in section 3 that in this
case the configuration of figure 2 is a local minimum of the classical energy and is separated
by a small potential bump from the true minimum, which is the configuration of figure 6 (the
energy landscape in this case is qualitatively depicted in figure 11). One can again choose the
parameters such that the whole process of tunneling takes place in the near-horizon region, and
only later is the information that it occurred communicated to the fivebranes. For these kinds
of non-perturbative instabilities one expects a similar picture to the one described above.

\item{(2)} Above we discussed cases in which the open string instability occurs deep inside
the near-horizon region of the fivebranes and the backreaction of the $NS'$-branes is a late
time effect. We expect the basic picture to hold in general. The open string instability
is always localized near the intersection of the various branes. Its resolution is the leading
dynamical effect, and once it occurs the bulk of the fivebranes reacts to the new
structure at the intersection in the way described above.

\item{(3)} The preceding discussion is very reminiscent of what happens in the process of localized
closed string tachyon condensation on non-compact orbifolds (see \eg\ \refs{\AdamsSV,\HarveyWM}).
Just like here the brane configuration can be characterized by a particular shape of the $NS5$-branes
at infinity, there the geometry is a cone of some particular opening angle infinitely far from the tip.
The analogs of open string instabilities in that case are tachyons localized at the tip (which come
from twisted sectors of the orbifold CFT). The asymptotic boundary conditions at infinity can change
under condensation of these tachyons, just like the asymptotic shape of the fivebranes in our examples.

\noindent
In the remainder of this section we will use the results of sections 3, 4 to propose a possible
form for the phase diagram of the model corresponding to the brane configurations of figures
2 -- 6. Our basic picture is summarized in figure 14 where we incorporate the distinction between
the cases\foot{As mentioned in subsection 3.4, it is possible that in the full open string theory,
beyond the DBI approximation, the critical value of $y$ is shifted from $l$ by $1/k$ corrections.}
$y>l$ and $y<l$ found in section 3.

\ifig\loc{The phase diagram of the intersecting brane model of figures 2 -- 6 for $y<l$ (a), and $y>l$ (b).
In figure 14b we use the notation $x^*\equiv \Delta x(\theta^*)$, $x_0\equiv \Delta x(\theta_0)$.}
{\epsfxsize4.2in\epsfbox{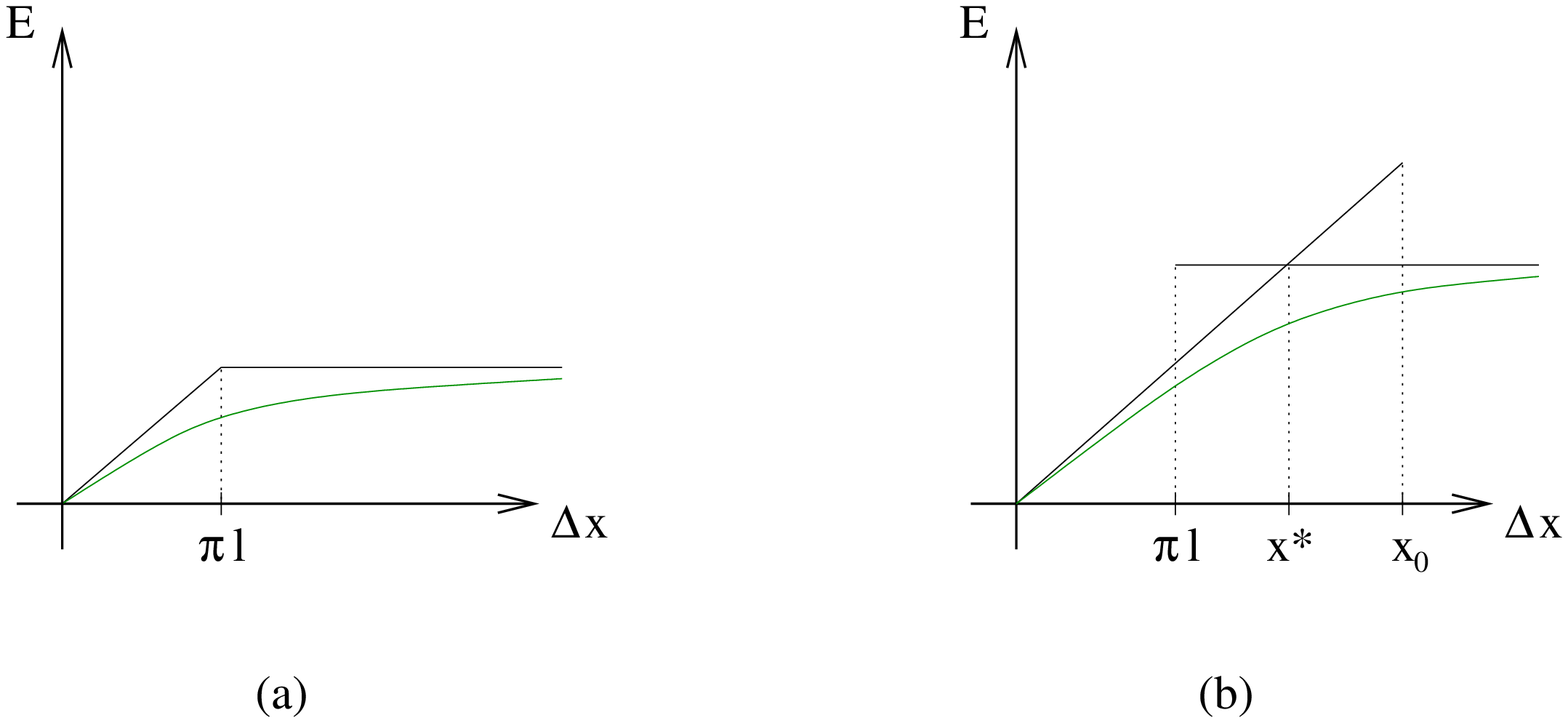}}

The horizontal black lines in figures 14a,b correspond to the brane configuration of figure 2. A result from
section 3 which is incorporated in the figure is that this configuration is a local minimum of the effective action
only for $\Delta x>\pi l$. The diagonal black lines in figures 14a,b correspond to the configuration of figure
6. For $\Delta x<\pi l$ this configuration is the only minimum of the effective action. For $\Delta x>\pi l$
the situation is different for $y<l$ and $y>l$. In the former case this solution ceases to exist in this regime.
Thus, the transition at $\Delta x=\pi l$ is second order, as explained in section 3 (see figure 9) and is
indicated in figure 14a. For $y>l$ this solution remains a local minimum of the effective action in a finite range
of $\Delta x>\pi l$, as indicated in figure 14b. The transition at $\Delta x=x^*$ is first order (see figure 13).

Both the horizontal and the diagonal black lines in figure 14, which 
were found in the classical analysis of section 3, describe 
non-supersymmetric brane configurations, so supersymmetry
is broken in this system, at least classically. An important
question is whether this is the case in the quantum theory as
well. The gauge theory analysis of section 4 suggests that the
answer is no. This analysis is valid in a small region near the
origin in figure 14, and it suggests that in that region there
exists a lower energy supersymmetric configuration, which we
denote in figure 14 by the curved magenta line. That line was
shown above to exist only in a small vicinity of the origin, but
it is natural to expect that it extends to all values of $\Delta
x$. It approaches one of the non-supersymmetric configurations
corresponding to the straight black lines in the two extreme
regions $\Delta x\to 0,\infty$.

As $\Delta x\to 0$ this can be understood from the gauge theory analysis of \IntriligatorDD\ and section 4.
In this region the mass parameter $m$ \formm\ is very small and the supersymmetric ground state becomes
almost indistinguishable from the meta-stable non-supersymmetric one. The latter is described in terms of branes
by the configuration of figures 3, 6 \refs{\OoguriBG\FrancoHT\BenaRG\AhnGN-\TatarDM}.

As $\Delta x\to\infty$, figure 14 suggests that the supersymmetric ground state approaches the brane configuration
of figure 2 where the separation between the $D4$ and $\overline{D4}$-branes goes to infinity. This is physically
reasonable -- while the configuration of figure 2 is not supersymmetric, in the limit $\Delta x\to\infty$ it
approaches one in which the dynamics of the $D4$-branes is decoupled from that of the $\overline{D4}$-branes.
The low energy theory for large $\Delta x$ is essentially the same for the non-supersymmetric brane
configuration of figure 2, and the supersymmetric one of figure 4a.  Both reduce to $N=1$ SYM with gauge group
$U(N_1)\times U(N_2)$ and no light matter.

We stress again that we have not proven that the supersymmetric ground state
described by the curved magenta line in figure 14 exists beyond the regime of small $\Delta x$
where one can establish its existence using gauge theory. It is not surprising that this is a
subtle problem. Indeed, translating the gauge theory analysis to string theory language, the
supersymmetric ground state should be highly quantum -- its existence should be due to
non-perturbative effects in $g_s$. Classically, such a vacuum does not exist. Furthermore, in
order to get to it in the field space of the brane theory, one has to turn on fields that are
non-geometric in the configurations of figures 2, 3, etc. Nevertheless, we expect that it should
be possible to find this ground state directly in string theory. The two main reasons for our belief in 
this are the following:
\item{(1)}The existence of a supersymmetric ground state at small $\Delta x$ follows from the 
field theory analysis and it would be surprising if it ceased to exist at a finite value of $\Delta x$. 
\item{(2)} The work on brane constructions in the late 1990's reviewed in \GiveonSR\ seems to suggest
that the low energy behavior behaves smoothly as one takes $NS'$-branes past $NS$-branes in configurations
such as that of figure 4. This is supposed to be the reflection of Seiberg duality in string theory
\ElitzurFH. Since in the electric brane configuration of figure 4 a supersymmetric ground state exists,
one would expect the same to be true in the magnetic one of figures 2, 3.

\noindent
A related point is that the phase structure we found is very similar
to what is expected in gauge theory \IntriligatorDD, despite the fact that
in our regime of parameter space the gauge theory analysis is not valid. In particular, in
\IntriligatorDD\ it was pointed out that as one increases the mass $m$ \formm, the meta-stable
state becomes less and less long-lived, and eventually it disappears for sufficiently
large $m$. Establishing this in gauge theory is not easy since it involves understanding the
theory in a regime where there is no weakly coupled description.

In our analysis we found precisely the same behavior in the DBI approximation (which, as mentioned
above, is valid in a different regime in the parameter space of brane configurations). Indeed, in
figure 14 the meta-stable state, which is described by the diagonal black line in figures 14a,b,
ceases to exist above a certain critical value. Moreover, the state it decays to has the same
qualitative features as the one in \IntriligatorDD. Unlike the gauge theory regime, here one can
analyze the system all the way to the point where the meta-stable state ceases to exist and beyond.
This is a common situation in string theory: in one regime understanding the dynamics involves
solving a non-trivial quantum field theory problem while in another one can analyze similar physics 
using classical string theory.

\newsec{Discussion}

In this paper we studied the phase structure of the brane system
of figures 2, 3 as a function of the parameters defining the brane
configuration. Most of our analysis was in the context of classical
string theory in the regime where all distances in figures 2, 3 are kept
finite in the limit $g_s\to 0$. We found that the system exhibits first
and second order phase transitions in different regions of its parameter space.

We also discussed a different region in parameter space, which can be
studied using an effective low energy gauge theory. In that region, we
saw (following \IntriligatorDD) that quantum effects lead to additional
vacua that preserve supersymmetry. The classical vacua become unstable,
but are parametrically long-lived in the limit $g_s\to 0$.

Much of the physics of the brane configuration we analyzed remains to
be understood. For example, it would be interesting to analyze the spectrum
of excitations of the configuration of figure 6. Classically, this
configuration is the ground state of the system for sufficiently small
$\Delta x$, and in some cases (for $y>l$) is locally stable for a range
of larger values of $\Delta x$ as well. Its low lying excitations can be
analyzed by expanding the DBI action about the solution \firstint, as is
familiar from studies of the Sakai-Sugimoto model \SakaiCN\ and other systems.

Perhaps the most interesting open problem is to understand whether the phase diagram of
figure 14 is correct, and the supersymmetric ground states that exist in gauge theory
persist for large $\Delta x$, where the gauge theory analysis is invalid. In gauge theory, 
the existence of these supersymmetric ground states is due to non-perturbative
effects. If the same is true in the stringy (large $\Delta x$) regime, it is important
to identify the relevant quantum effects and provide a good description of the
supersymmetric vacua.

There are many natural generalizations of the brane configurations studied here.
For example, replacing the $NS'_2$-brane by $N_2$ $D6$-branes stretched in $(0123789)$ 
leads to the brane configuration corresponding to the magnetic dual of supersymmetric
QCD with gauge group $U(N_1)$ and $N_2$ chiral superfields in the fundamental representation
\ElitzurFH. The separation of the $NS'_1$-brane and $D6$-branes in the $x$ direction, 
that in gauge theory leads to meta-stable non-supersymmetric vacua \IntriligatorDD, was 
discussed using branes in \refs{\OoguriBG\FrancoHT\BenaRG\AhnGN-\TatarDM}.

Our DBI analysis is applicable to this case as well since it does not depend on what the
$D4$-branes are ending on. In fact, this case is in some ways better behaved since the 
quantum brane bending effects \largewbend\ are smaller. In particular, the bending of
the $D6$-branes due to the $D4$-branes ending on them goes to zero asymptotically. The
classical string theory analysis of section 3 leads in this case to a picture
very reminiscent of that of \IntriligatorDD. The phase diagram is again expected to be given by
figure 14. The origin of the supersymmetric ground state is again not obvious, and it would be
interesting to understand it better.

One can also consider the system  in which the $NS'_1$-brane is
replaced by $D6$-branes as well. The DBI analysis is still the
same,  but in this case there is no reason to expect 
non-trivial quantum dynamics to restore supersymmetry. Indeed, the
low energy theory on the $D4$-branes is in this case weakly
coupled, and there are no large bending effects of the sort
\largewbend, since there are no longer any $NS'$-branes.\foot{The
$NS$-branes play a very different role in our construction. For
example, for $N_1=N_2$ the total charge on them is zero, so
asymptotically, for large $(x^4, x^5)$, they retain their
classical shape.}

Many other generalizations are suggested by the work on
supersymmetric brane constructions reviewed in \GiveonSR. For
example, the original work of \HananyIE\ involved $D3$-branes
(rather than $D4$-branes) suspended between $NS5$-branes, and led to
new insights into $2+1$ dimensional gauge dynamics. It is natural
to consider similar generalization in non-supersymmetric setups
such as those studied in this paper.

As mentioned in the introduction, a class of related constructions
involves branes and antibranes wrapping different cycles of non-compact
CY manifolds. An example analyzed in \refs{\AganagicEX,\HeckmanWK} is
the surface 
\eqn\twocon{y^2=W'(x)^2+uv~,}
with
\eqn\formsup{W'(x)=(x-a_1)(x-a_2)~.}
This non-compact manifold can be thought of as two adjacent conifold 
singularities located at $x=a_1, a_2$. One can resolve each conifold 
by blowing up certain two-spheres in the geometry, and wrap $D5$
and $\overline{D5}$-branes (which are also stretched in the usual 
Minkowski spacetime $\IR^{3,1}$), respectively, around them.  

The non-compact CY \twocon, \formsup\ is related by T-duality 
\refs{\OoguriWJ\KutasovTE-\GiveonZM} to a background which
contains an $NS5$-brane wrapped around the surface 
\eqn\wrappedfive{y^2=W'(x)^2~.}
This is actually two fivebranes wrapped around the surfaces $y=\pm W'(x)$
in the $\IC^2$ labeled by the complex coordinates $(x,y)$. The ten 
dimensional target space of type II string theory contains this $\IC^2$, the 
physical Minkowski spacetime, $\IR^{3,1}$, and two additional dimensions, 
in which the two fivebranes can be separated. The $D5$-branes 
and antibranes of \refs{\AganagicEX,\HeckmanWK} correspond in this description 
to $D4$-branes and antibranes that are stretched between the two $NS5$-branes, 
the $D4$-branes at $x=a_1$ and the $\overline{D4}$-branes at $x=a_2$.  

From the perspective of our discussion, this system is simpler to understand 
than the one of figure 2. The branes and antibranes are locally stable, as in our
case, but they can still annihilate by first increasing their energy. For example,
if the number of $D4$ and $\overline{D4}$-branes are equal, their ends on one of
the $NS5$-branes can approach each other and reconnect, such that the $D$-branes 
turn into a single stack both of whose ends lie on the other fivebrane. They can
then shrink to zero size and disappear. 

Thus, some of the non-trivial aspects of the analysis above are absent in
this case. The classical ground state of the model is the supersymmetric state 
with no $D$-branes, as opposed to our case where it was one of the non-supersymmetric 
configurations of figures 2 or 6, depending on the parameters. The semiclassical 
phase diagram is simpler, and in particular the phase transitions we found are absent. 
Nevertheless, one can hope that some of the techniques that were used to study that
system can shed additional light on the one considered here and generalizations thereof.

\bigskip\bigskip
\noindent{\bf Acknowledgements:} We thank A. Parnachev for 
discussions. This work is supported in part by
the BSF -- American-Israel Bi-National Science Foundation. The
work of AG is supported in part by the center of excellence
supported by the Israel Science Foundation (grant No. 1468/06), EU
grant MRTN-CT-2004-512194, the DIP grant H.52, and the Einstein
Center at the Hebrew University. The work of DK is supported in
part by DOE grant DE-FG02-90ER40560 and the National Science
Foundation under Grant 0529954. AG thanks the EFI at the
University of Chicago for hospitality. DK thanks the Weizmann
Institute and Hebrew University.

\listrefs
\end